\documentclass[nonacm]{acmart}

\usepackage{multirow}
\usepackage{enumitem}
\usepackage{tabularx}

\copyrightyear{2026}
\acmYear{2026}
\setcopyright{cc}
\setcctype{by}
\acmConference[CHI '26]{Proceedings of the 2026 CHI Conference on Human Factors in Computing Systems}{April 13--17, 2026}{Barcelona, Spain}
\acmBooktitle{Proceedings of the 2026 CHI Conference on Human Factors in Computing Systems (CHI '26), April 13--17, 2026, Barcelona, Spain}
\acmDOI{10.1145/3772318.3791347}
\acmISBN{979-8-4007-2278-3/2026/04}

\begin{document}

\title[How Machine Learning Practitioners at Big Tech Companies Approach Fairness in Recommender Systems]{Fairness-in-the-Workflow: How Machine Learning Practitioners at Big Tech Companies Approach Fairness in Recommender Systems}

\author{Jing Nathan Yan}\authornote{Co-first authors.}
\email{jy858@cornell.edu}
\affiliation{%
  \institution{Cornell Tech}
  \city{New York}
  \state{NY}
  \country{USA}
}

\author{Emma Harvey}\authornotemark[1]
\email{evh29@cornell.edu}
\orcid{0000-0001-8453-4963}
\affiliation{%
  \institution{Cornell Tech}
  \city{New York}
  \state{NY}
  \country{USA}
}

\author{Junxiong Wang}
\email{jw2544@cornell.edu}
\affiliation{%
  \institution{Cornell University}
  \city{Ithaca}
  \state{NY}
  \country{USA}
}

\author{Jeffrey M. Rzeszotarski}
\email{jeff.rzeszotarski@gmail.com}
\affiliation{%
  \institution{Loyola University Maryland}
  \city{Baltimore}
  \state{MD}
  \country{USA}
}

\author{Allison Koenecke}
\email{koenecke@cornell.edu}
\affiliation{%
  \institution{Cornell Tech}
  \city{New York}
  \state{NY}
  \country{USA}
}

\renewcommand{\shortauthors}{Yan and Harvey et al.}

\begin{abstract}
Recommender systems (RS), which are widely deployed across high-stakes domains, are susceptible to biases that can cause large-scale societal impacts. Researchers have proposed methods to measure and mitigate such biases---but translating academic theory into practice is inherently challenging. Through a semi-structured interview study (N=11), we map the RS practitioner workflow within large technology companies, focusing on how technical teams consider fairness internally and in collaboration with legal, data, and fairness teams. We identify key challenges to incorporating fairness into existing RS workflows: defining fairness in RS contexts, balancing multi-stakeholder interests, and navigating dynamic environments. We also identify key organization-wide challenges: making time for fairness work and facilitating cross-team communication. Finally, we offer actionable recommendations for the RS community, including practitioners and HCI researchers.
\end{abstract}



\maketitle

\section{Introduction}
Recommender systems (RS) are inherently social and collaborative entities~\cite{stray2022building, li_beyond_2025} which connect providers (e.g., sellers, content creators) with users (e.g., buyers, content consumers). RS are widely deployed across high-stakes domains including e-commerce~\cite{gretzel2006persuasion, zalmout2021all} and social media~\cite{lada2021machine, belli2020privacy, anandhan2018social}---but they are susceptible to biases that can cause large-scale psychological, ideological, and societal impacts~\cite{li2021tutorial, chen2020bias, kay2015unequal, sweeney_discrimination_2013}. For example, Amazon's e-commerce RS shows preference to Amazon's own products even when competing products have superior reviews, resulting in an effective monopoly~\cite{markupAmazon}, and X's social media RS disproportionately amplifies right-leaning political content, resulting in potential electoral impacts~\cite{ye2025auditing, huszar2022algorithmic}. ML, HCI, and policy researchers have proposed a variety of methods to measure and mitigate such biases in order to improve fairness in RS~\cite{feng_has_2022, koenecke2023popular, smith2024recommend, abdollahpouri2019managing}. However, appropriately translating any academic theory into industry application is inherently challenging because it requires navigating real-world constraints and balancing conflicting stakeholder interests \cite{holstein2019improving, deng2022exploring, madaio2020co, beutel2019putting, veale2018fairness, metcalf_owning_2019, robertson_not_2022, lee_landscape_2021, quinonero_candela_disentangling_2023, rieke_imperfect_2022, smith2022real}. Incorporating fairness considerations into RS development is particularly challenging because RS operate within a landscape where fairness is not a static concept but a dynamic one, with meanings that shift across different contexts~\cite{smith2024recommend, smith_pragmatic_2025, ekstrand2022fairness, smith2023many}.

The complexity of RS is amplified by its operation in a multi-sided market, involving diverse stakeholders---both providers and end-users, each with heterogeneous make-ups~\cite{burke2017multisided, abdollahpouri2019multi}. RS practitioners who seek to incorporate fairness into their workflows therefore face the unique challenge of balancing multiple competing sets of stakeholder interests. Furthermore, RS practitioners operate subject to multiple ethical and regulatory constraints, such as the need to ensure the dissemination of trustworthy information, and sometimes must make tradeoffs between prioritizing fairness and those other considerations. RS practitioners also do not operate in a static environment. Their roles require not only identifying existing sources of unfairness, but anticipating emerging ones and mitigating their long-term effects, which can be exacerbated by user feedback loops~\cite{burke2017multisided, li2021tutorial, li_beyond_2025}. This task demands a high level of foresight and adaptability---practitioners must be proactive in their approach to improving fairness in RS.

The ubiquity of RS today underscores the critical role that RS practitioners play in shaping society's online interactions. Technical RS teams serve as the central hub where innovation, ethical considerations, and practical applications intersect. Here, the core challenges of fairness in RS are actively addressed and navigated. Hence, building on existing research, we explore the unique nuances of how machine learning (ML) practitioners at `big tech' companies approach fairness in RS. In particular, \textbf{we map the RS practitioner workflow, with an emphasis on how technical teams attempt to incorporate fairness (a) into their existing internal practices and (b) in collaboration with legal, data, and fairness teams.} This allows us to understand how technical practitioners approach fairness in RS---a focus that is critical in understanding how to bridge academic theory with actual practices. To do this, \textbf{we conduct semi-structured interviews with technical RS practitioners who are integral members of ML teams at big tech companies} (N=11).  We examine what they have done to integrate fairness into their workflows, as well as the technical and organizational challenges they have faced to doing so. Technically, practitioners struggle to define fairness in the context of RS, especially because they must account for complexities inherent to RS, which include considering fairness across large numbers of (conflicting) stakeholder groups and working in dynamic settings with deployed systems that are susceptible to feedback loops. Organizationally, practitioners struggle with a lack of time to conduct fairness work and with developing a `fairness lingua franca' that allows for clear communication across multidisciplinary teams. Our findings reveal opportunities for practitioners and HCI researchers to improve fairness in RS. Our key contributions are:
\begin{itemize}
  \item A map of the RS workflow, emphasizing how technical teams approach fairness (a) internally and (b) through collaborations with legal, data, and fairness teams
  \item An overview of the key technical and organizational challenges to incorporating fairness into the RS workflow
  \item A set of actionable recommendations to better integrate fairness into RS workflows, highlighting both where big tech companies and practitioners are responsible for implementing evidence-based changes as well as where novel HCI research can support real-world fairness efforts
  \end{itemize}

We proceed as follows. First, we provide an overview of prior work on fairness considerations for RS and on the challenges inherent to translating academic theory into practice (\S\ref{sec:rw}). We describe our semi-structured interview and analysis approach (\S\ref{sec:method}). Then, we provide an overview of the RS workflow (\S\ref{sec:workflow}). We describe technical challenges (\S\ref{sec:thematic}) and organizational challenges (\S\ref{sec:org}) to implementing fairness in RS workflows. We make calls to action for RS practitioners and HCI researchers (\S\ref{sec:calls}) and close with a discussion of the limitations of our work and potential paths forward (\S\ref{sec:discussion}).
\section{Related Work}\label{sec:rw}

Here, we provide a brief overview of the academic literature on ML fairness, with a focus on fairness for RS. We also discuss prior work on implementing fairness in deployed systems, with a focus on the challenges inherent to translating academic principles into practice.

\subsection{Fairness in Machine Learning}\label{subsec:background:fairML}

Fairness and bias are closely related but distinct. \textit{Bias} refers to systematic errors or deviations in data or models that produce skewed outcomes for certain groups~\cite{hutchinson201950}. \textit{Fairness}, by contrast, concerns the ethical acceptability of those biased outcomes~\cite{corbett2023measure}. We focus on fairness in this paper.\footnote{We note that during interviews, participants sometimes used the terms `bias' to refer to what we generally call `unfairness,' or `debiasing' to refer to what we generally call `improving fairness,' so we use those terms when quoting participants directly.} Fairness is an essentially contested concept, meaning that there is no single agreed-upon approach for defining fairness~\cite{jacobs_measurement_2021}. A dominant fairness paradigm focuses on the outcomes of technical systems: systems are \textit{group fair} to the extent that they produce similar outcomes for different groups of individuals, and \textit{individually fair} to the extent that they treat `similar' individuals similarly \cite{dwork2012fairness, barocas2016big}. Although these approaches sound simple, they are anything but: group fairness can be defined using multiple metrics, many of which are impossible to simultaneously satisfy, and individual fairness requires a way of determining how similar any two individuals are, which is often infeasible in practice \cite{chouldechova2017fair, kleinberg2016inherent, zemel2013learning}. An alternative paradigm draws on the theory of procedural justice \cite{rawls_theory_1999}, arguing that systems are fair to the extent that they follow fair processes---although what makes a process fair is not settled, and may involve anything from only using certain features~\cite{grgic-hlaca_case_2016} to incorporating multiple stakeholder groups in participatory system development~\cite{lee_webuildai_2019}. Finally, fairness is but one component of the broader field of ethical or responsible AI. Thus, the choice of how to define fairness often intersects with other considerations. These may include legal obligations (e.g., anti-discrimination laws)~\cite{hildebrandt_issue_2022, Magrani2024}, explainability~\cite{kunkel_let_2019}, user privacy considerations~\cite{sun_when_2023}, and the extent to which systems facilitate appeal and recourse of algorithmic decisions~\cite{jin_beyond_2024}. Ultimately, organizations must choose which principles to prioritize, and the choice is far from simple~\cite{hutchinson201950,ekstrand2022fairness}. 

\paragraph{\textbf{Fairness for Recommender Systems}}
Defining fairness is particularly challenging in the context of RS because much ML fairness research has focused on classification problems, meaning that many existing fairness objectives and metrics may not be directly applicable to RS. For example, although fairness in classification problems often focuses on outcomes, what an outcome means in the context of RS is not necessarily clear: outcomes can be thought of in terms of not only \textit{allocations} (e.g., whose content gets recommended) but also \textit{representations} (e.g., whether recommended content contributes to the stereotyping, erasure, or demeaning of social groups)~\cite{blodgett_language_2020}. Nevertheless, researchers have proposed a variety of approaches for measuring and improving fairness in recommendations and ranking~\cite{Zehlike_fair_2017, asudeh_designing_2019, Biega_equity_2018, ali_discrimination_2019, beutel_fairness_2019, diaz_evaluating_2020, pmlr-v81-ekstrand18b, geyik_fairness_2019, li_user_2021, zehlike_reducing_2020, singh_fairness_2018, steck_calibrated_2018, ye_measuring_2017, rastogi_fairness_2024}.\footnote{For comprehensive overviews of fairness in RS, we direct the reader to \citet{ekstrand2022fairness}, \citet{patro_fair_2022}, and \citet{zehlike_fairness_2022a, zehlike_fairness_2022b}. For a survey of fair ranking metrics, we point to \citet{smith2023scoping}.} As discussed above, defining fairness (e.g., by choosing a fairness metric) is challenging in ML generally. It is uniquely challenging in the context of RS because RS require balancing the (potentially competing) interests of multiple stakeholders who are dynamically interacting with each other and the RS platform at massive scales~\cite{chen2020bias, wang2022trustworthy, abdollahpouri2019multi,  burke2017multisided}. In particular, RS practitioners must consider the interests of \textit{providers} (e.g., sellers) and \textit{end-users} (e.g., buyers) ~\cite{burke2018balanced,singh_fairness_2018,sonboli2021fairness, leonhardt2018user, li_user_2021, stray2022building, mehrotra2018towards}. For instance, a user-centric fairness approach might target equally accurate recommendations across demographics~\cite{sonboli2021fairness}, while a provider-focused approach might seek equitable exposure for underrepresented creators~\cite{abdollahpouri2019managing}---and these two interests can easily be in direct conflict with one another~\cite{hutchinson201950,sonboli2021fairness, ekstrand2022fairness,smith2023many}. A system that prioritizes promoting content from historically marginalized providers might inadvertently serve users content that is misaligned with their previously stated interests~\cite{smith2023many}. In fact, recent user studies have found that users often prefer gender-biased recommendations over `debiased' ones, even when they are made aware that bias is present~\cite{wang_biased_2023, krause_effect_2025}. As \citet{smith2024recommend} point out, a `debiased' system can even have potentially negative effects for providers by exposing them to harassment from users who are outside of their preferred target audience. Complicating things further, a single platform may contain multiple types of providers, such as social media content creators and paid advertisers, whose interests may conflict as well. Even within a stakeholder group, how fairness should be defined is not straightforward. For example, when considering providers in a video-sharing platform, while one approach to fairness might involve rewarding providers for producing high-quality content by using straightforward viewer metrics for ranking videos, an alternative approach might incorporate randomness in search rankings to counteract the ``rich-get-richer'' effect~\cite{chen2020bias} wherein top-ranked videos get disproportionately more future views.

The massive scale of RS also raises unique challenges in practice. As \citet{anthis-etal-2025-impossibility} argue, technologies that are used by large populations across many contexts, like RS, simply cannot be made fair for every context in which they operate and every population subgroup with which they interact. Scale also causes technical challenges related to sparse user-content interaction data, because users may only interact with a tiny subset of overall content~\cite{li_user_2021, kearns2019empirical,beutel2019putting, beutel_fairness_2019, beutel2018latent,lum2022biasing}. Finally, as RS use user-content interaction data generated from the recommendation process to inform their future results, they risk feedback loops that amplify disparities~\cite{keller2021amplification, li_beyond_2025}. Feedback loops themselves are multi-faceted: users can provide \textit{implicit} feedback (e.g., quickly scrolling past a video) or \textit{explicit} feedback (e.g., requesting not to see a video again) \cite{oard1998implicit, li_beyond_2025}. Users provide explicit feedback less consistently than they provide implicit feedback \cite{lustig_designing_2022}; however, implicit feedback is harder to interpret and is often itself shaped by users' folk theories of how the RS they interact with work \cite{eslami_first_2016, devito_algorithms_2017, xiao_influence_2025}. The choices by RS practitioners about what feedback to include---and the choices by users about how to convey and organize their feedback---can thus have major impacts on fairness in RS. \textit{Our study builds on the fair ML literature by examining how inherent fairness tensions are approached in practice and the extent to which academic definitions of fairness guide industry approaches, taking the unique challenges of RS fairness into account.}

\subsection{Fairness in Practice}

\paragraph{\textbf{Bringing About Fairness Efforts}} In practice, fairness efforts do not occur in isolation. Rather, they interact---and may come into tension---with organizational priorities, such as maximizing profits. Acquiring organizational support to conduct responsible AI initiatives often requires linking fairness to explicit business or product goals~\cite{scheuerman2024walled, ali2023walking, garcia-gathright_assessing_2018, rakova2021responsible, veale2018fairness, wang_strategies_2024}. Regulation can also play a role in bringing about responsible AI efforts. However, regulations often vary across legal jurisdictions, complicating fairness efforts for multinational platforms that use RS. For example, RS in the EU are subject to the Digital Services Act (DSA), which authorizes yearly assessments of online platforms regarding fairness issues, illegal content, and disruption to human rights~\cite{madiega2020digital}. The DSA does not apply to RS in the US \cite{wired_dsa}; however, individual states have created their own fairness regulations, such as California's 2022 state law AB 587,\footnote{\url{https://openstates.org/ca/bills/20212022/AB587/}} which aims to provide transparency about large social media RS to avoid disinformation and hate speech. When fairness standards are not mandated by law, organizations may avoid assessing fairness altogether due to concerns that uncovering unfairness could add to organizational liability~\cite{costanza-chock_who_2022, wright_null_2024}. If institutional support is lacking, employees may choose to ``individualize risk''~\cite{ali2023walking} by taking personal responsibility for incorporating fairness into their workflows~\cite{ganesh_wild_2025}.

\paragraph{\textbf{Operationalizing Fairness}} However efforts are initiated, doing fairness in practice requires adhering to a specific operationalization of fairness. For example, documentation efforts like datasheets for datasets \cite{gebru_datasheets_2021}, model cards for model reporting \cite{mitchell_model_2019}, and FactSheets \cite{arnold2019factsheetsincreasingtrustai} call on practitioners to publicly share standardized reports describing the intended use, disaggregated performance, and other fairness-related characteristics of their data and models. Similarly, fairness checklists, which are inspired by safety checklists used in medicine and aviation, are intended to encourage practitioners to consider fairness issues and mitigate potential sources of unfairness throughout the design and development of technical systems~\cite{madaio2020co}. As a more technical operationalization, much fair ML research focuses on developing metrics and toolkits that practitioners can use to automatically measure and improve fairness in decision-making systems~\cite{ojewale_2025_towards, wong_seeing_2023}.

\paragraph{\textbf{Fairness Challenges in Practice}} Prior work has repeatedly shown that efforts to operationalize fairness within organizations are deeply challenging. Practitioners struggle to reconcile incompatible fairness metrics~\cite{madaio2022assessing, moharana2025accessibilitypeopleworkthing, smith_pragmatic_2025,koenecke2023popular}, work with messy real-world data~\cite{cramer_translation_2019, holstein2019improving, rieke_imperfect_2022, guerdan2025measurementbricolageexaminingdata, passi2019problem, sambasivan_everyone_2021,pang2023auditing}, and adapt general-purpose academic tools and metrics to their specific real-world needs~\cite{deng2022exploring, harvey-etal-2025-understanding, harvey2024gapsresearchpracticemeasuring, lee_landscape_2021, balayn__2023, kaur_interpreting_2020, richardson_towards_2021, ojewale_2025_towards, madaio_tinker_2024, Morley2020,Corpus2025}. Challenges are exacerbated by organizational barriers to communicating and collaborating across different teams and within cross-functional teams \cite{deng2023investigating}, or to incorporating the perspectives of multiple stakeholders, including end users, into fairness assessments and improvements \cite{deng_understanding_2023, costanza-chock_who_2022, madaio2020co}.

With the notable exception of Smith et al. \cite{smith2023many, smith2023scoping, smith2024recommend, smith_pragmatic_2025}, most prior work on implementing fairness in practice has focused broadly on  general-purpose ML or specifically on fairness in classification models, NLP models, or other non-RS applications of ML. However, we expect the issues raised by prior work to be even more pressing in the RS space, where fewer academic toolkits are available to scaffold fairness efforts, and where the interests of multiple stakeholders (providers and users) may be in direct competition. In fact, Smith et al.~have previously identified tensions between competing stakeholders through interview studies with individuals who interact with RS but do not work directly on technical RS teams, including providers (content creators and dating app users)~\cite{smith2024recommend} and ML fairness experts~\cite{smith_pragmatic_2025}. Most similar to our work, Smith et al.~have conducted interview studies with RS practitioners themselves: employees of an anonymous ``commercial consumer-provider recommender system''~\cite{smith2023scoping} and of a microlending platform~\cite{smith2023many}. Both studies explored challenges related to selecting fairness metrics in the context of RS, again identifying competing stakeholder interests as a key challenge \cite{smith2023many, smith2023scoping}. In contrast, we interview technical RS practitioners across multiple big tech companies, and look beyond fairness metrics to how fairness is incorporated into RS workflows more broadly. \textit{In summary, we build on the extensive body of prior work by focusing explicitly on the RS workflow, with a spotlight on how RS practitioners across multiple big tech companies navigate not only technical challenges stemming from RS-specific complexity, but also organizational challenges stemming from cross-team interactions with legal, data, and fairness practitioners.}
\section{Methods}\label{sec:method}

To understand how RS practitioners approach fairness in practice, we conducted 11 semi-structured interviews over a three-week period in 2022. We interviewed RS practitioners from a variety of large companies that deploy highly visible RS. Our interviews were intended to answer two key research questions: 
\begin{itemize}
\item[\textbf{RQ1:}] What do the technical and fairness workflows of an RS practitioner look like, and how do they intersect? 
\item[\textbf{RQ2:}] What challenges do RS practitioners face when incorporating fairness considerations into their technical workflows? 
\end{itemize}

\paragraph{\textbf{Recruitment and Participants}}
We recruited participants via social media, direct emails to individuals in our networks or who we had identified based on publicly available information as working in RS, and snowball sampling. Participation was voluntary, and all participants provided informed consent prior to the start of interviews. Following the interviews, participants were compensated with Amazon gift cards. All interviews were conducted via video chat. Participants were given the option of either having interviews recorded to create audio transcripts, or having researchers take detailed notes during interviews. We provided the option for note-taking instead of recording as some practitioners expressed concerns about accidentally disclosing confidential company information, although no such disclosures ultimately occurred. In total, we recruited 11 ML practitioners (averaging 5.36 years of experience) across seven big tech companies. Practitioners held titles such as `Machine Learning Engineer,' `Machine Learning Researcher,' or `Scientist,' and were actively involved in building RS. We specifically focused on technical practitioners because we were interested in understanding how fairness considerations are integrated into the work of building RS and how practitioners handle dual technical and fairness responsibilities. (We discuss the implications of this decision in more detail in \S\ref{subsect:limitations}.) Each participant worked for a company that serves over one billion customers annually. Further details on participants are available in Table~\ref{tab:participant}. This study was supervised by a university Institutional Review Board (IRB).

\begin{table*}[ht]\small
\begin{tabular}{cccc}
\toprule
\textbf{Participant ID} & \textbf{Years of Experience} & \textbf{Engineering Responsibilities} & \textbf{Research Responsibilities}  \\ \midrule
P1  & 3 & \checkmark & \checkmark \\ 
P2  & 4 & \checkmark & \checkmark \\ 
P3  & 4 & \checkmark & \checkmark \\ 
P4  & 5 & \checkmark & \checkmark \\ 
P5  & 6 & \checkmark & \\
P6  & 5 & \checkmark & \checkmark \\ 
P7  & 4 & \checkmark & \\
P8  & 5 & \checkmark & \checkmark \\
P9  & 7 & \checkmark & \checkmark \\ 
P10 & 3 & \checkmark & \checkmark \\
P11 & 12 & \checkmark & \checkmark \\ \bottomrule
\end{tabular}
\caption{\textbf{Participant Background Information.}}\label{tab:participant}
\Description{A table describing participant background. There are 11 participants (P1-P11). All participants held roles that involved engineering responsibilities, and all but P5 and P7 held roles that involved research responsibilities. Participants had varying years of experience: 3, 4, 4, 5, 6, 5, 4, 5, 7, 3, and 12, respectively.}
\end{table*}

\paragraph{\textbf{Interviews}} Interviews ranged between 45 minutes and one hour. Practitioners were instructed to be as detailed as possible but to avoid sharing any confidential information such as revenue or upcoming products. We began by asking ice-breaking questions regarding practitioner background, including their years of experience, employer, and the RS that they built or worked with. Next, we asked questions intended to help us map practitioner workflows (RQ1). We began with open-ended questions (e.g., ``could you describe a recent project's workflow?'', ``are there external teams that you interact with regularly?'') and asked more detailed follow-up questions if necessary (e.g., ``how do you feel when interacting with the data team?''). We chose to begin with these open-ended, workflow-centric questions in order to develop an understanding of how central fairness work was to practitioners' broad descriptions of their responsibilities before they were explicitly prompted to consider fairness. Next, we asked about participants' professional experience with assessing and improving fairness in RS. These included workflow-related questions intended to address RQ1 (e.g., ``what amount of time do you spend on fairness in your daily work?'') as well as questions related to potential challenges intended to address RQ2 (e.g., ``can you describe a recent time that improving fairness was difficult or failed?''). We also asked about participants' perceptions of their organizations' approaches to fairness (e.g., ``is fairness in your organization a team effort?''). We asked about these separately to try to disentagle participants' technical considerations and challenges from organizational ones. Interview quotes included in this paper have been lightly edited for clarity. Our semi-structured interview guide is provided in Appendix Table \ref{tab:example:more_question}.

\paragraph{\textbf{Analysis}}
We sought to understand how fairness is approached in practice by individuals and teams who develop and deploy large-scale RS. To do so, we analyzed responses on two levels: first, we mapped RS practitioner workflows, and second, we conducted an inductive thematic analysis to understand the challenges faced by RS practitioners who seek to improve fairness in the systems that they build, deploy, and maintain. 

\textit{RQ1: Mapping Practitioner Workflows.} First, we developed a unified mapping of practitioners' workflows. As our goal was to recognize general patterns among practitioners, we focused on identifying commonalities and flagged outlier cases for discussion and further examination. Data were analyzed during collaborative meetings between authors, with interviewers summarizing their sessions and outlining transcripts and/or notes relevant to each session under inspection. As each transcript or set of notes was discussed, we identified key stages to building an RS that were discussed by practitioners. We also identified the core set of external teams with whom practitioners interacted. We then collaboratively sketched and refined workflow models describing both the general flow of data during model development as well as the specific fairness approaches undertaken by practitioners at each stage of the process. Findings are described in \S\ref{sec:workflow}.

\textit{RQ2: Understanding Practitioner Challenges.} Second, we conducted an inductive thematic analysis to identify core challenges faced by practitioners as they sought to incorporate fairness considerations into their technical workflow \cite{braun_using_2006}. We first converted transcripts and interview notes into items on a Miro web board for aggregation and analysis. Authors conducted two coding passes through all quotes and notes, developing a large set of initial codes related to any mention of challenges that practitioners faced when trying to incorporate fairness into their technical RS workflows. Next, authors collated codes into initial themes. Then, authors reviewed and refined themes, aggregating thematic groups, merging similar themes, and recategorizing elements through negotiation. Lastly, through another open discussion, authors broke down themes that had a significantly larger number of notes into finer granularity and placed any remaining unlabeled items into categories. When finalizing results, authors summarized the high-level themes identified and pulled representative quotes from the grouped items. Findings about technical challenges faced by RS practitioners are described in \S\ref{sec:thematic}, and findings about organizational challenges arising from RS practitioner interactions with other teams are described in \S\ref{sec:org}.

\paragraph{\textbf{Positionality}}
We are a team of researchers who work on academic and industry teams, and who collectively have expertise in machine learning, human-computer interaction, and algorithmic fairness. Our backgrounds have informed our approach to this study. Our experience on industry teams informed our understanding of the practical constraints that practitioners may face, which influenced the questions we asked in our semi-structured interviews. At the same time, our experience does not extend to working as RS practitioners within technology companies. While this positions us to critically assess RS workflows and challenges as outsiders, we lack `insider' knowledge about what it is like to work on RS in practice. Finally, we are also all end-users of RS systems. This dual vantage point as both researchers and users shaped our understanding of the potential real-world impacts of RS, which in turn informed our approach to thematic analysis.
\section{The RS Practitioner Workflow}\label{sec:workflow}

\begin{figure*}[t]
\centering
\includegraphics[width=\textwidth]{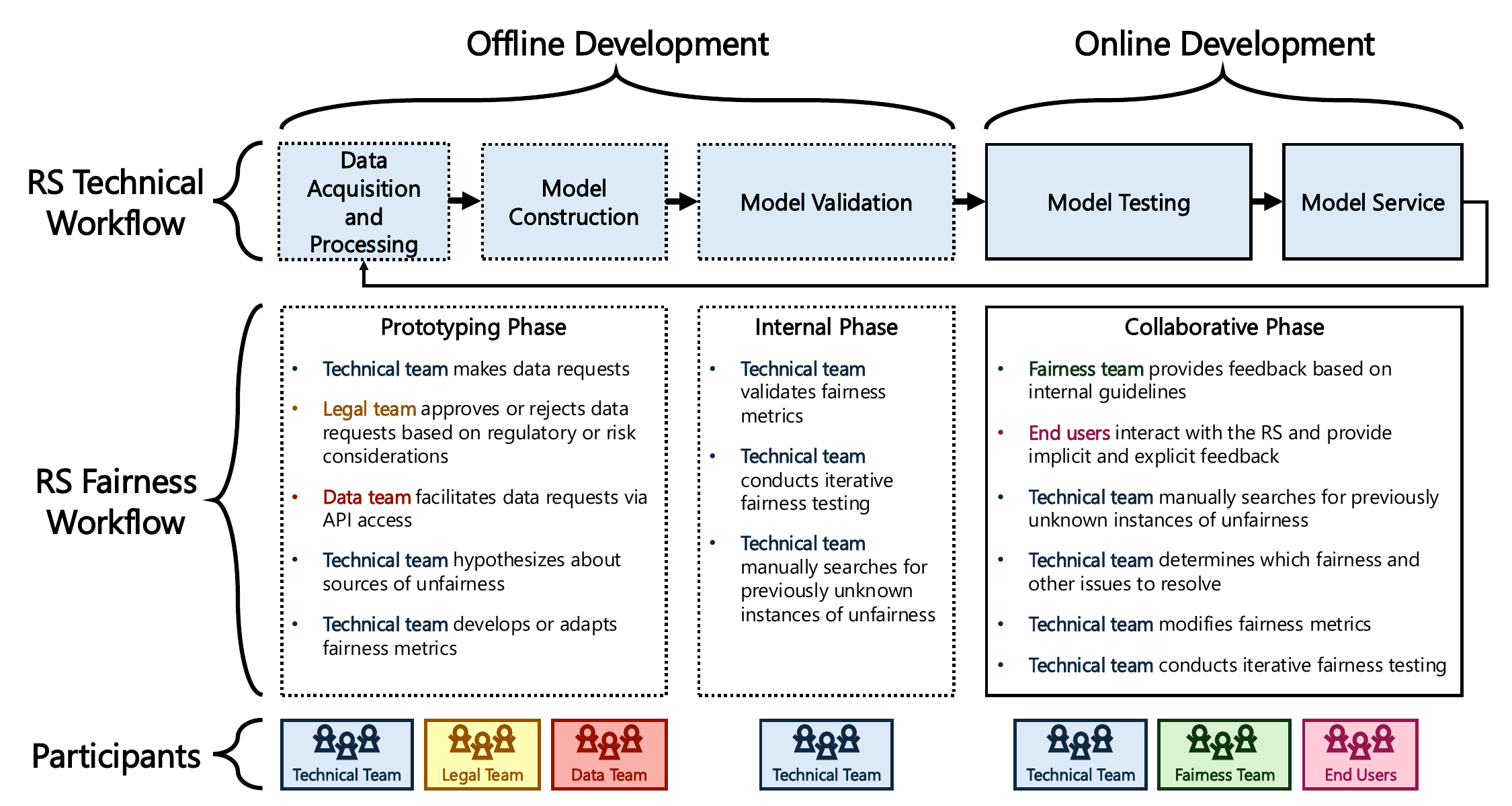}
\caption{
\textbf{Recommender System Practitioner Workflow.} We visualize RS practitioners' interactions with the data team managing and monitoring production data access, the legal team auditing and approving specific data access requests, and the fairness (or responsible AI) team providing fairness expertise. Dotted borders indicate offline phases of the workflow, and solid borders indicate online phases. For each phase of the technical workflow (top row), there are associated phases within the fairness workflow (second row). In each phase, different participants interact (bottom row).\looseness=-1}\label{fig:workflow}
\Description{A figure summarizing the RS practitioner technical and fairness workflows. The technical workflow includes the following offline phases: data acquisition and processing, model construction, and model validation; as well as the following online phases: model testing and model service. The fairness workflow co-occurs with the technical workflow. The fairness workflow consists of the prototyping, internal, and collaborative phases. The prototyping phase, which co-occurs with the offline data acquisition and processing and model construction phases, includes the technical team, the legal team, and the data team. In the protoyping phase, the following actions are taken: the technical team makes data requests, the legal team approves or rejects data requests based on regulatory or risk considerations, the data team facilitates data requests via API access, the technical team hypothesizes about sources of unfairness, and the technical team develops or adapts fairness metrics. The internal phase, which co-occurs with the offline model validation phase, includes the technical team. In the internal phase, the technical team validates fairness metrics, conducts iterative fairness testing, and manually searches for previously unknown instances of unfairness. The collaborative phase, which co-occurs with the online model testing and model service phases, includes the technical team, the fairness team, and end users. In the collaborative phase, the fairness team provides feedback based on internal guidelines; end users interact with the RS and provide implicit and explicit feedback; and the technical team manually searches for previously unknown instances of unfairness, determines which fairness and other issues to resolve, modifies fairness metrics, and conducts iterative fairness testing.}
\end{figure*}

In this section, we describe the RS practitioner workflow. We focus on how RS practitioners integrate fairness work internally (within technical teams) and in collaboration with legal, data, and fairness teams.

\subsection{An Overview of the RS Practitioner Workflow}  
The RS practitioner workflow is summarized in Figure~\ref{fig:workflow}. Practitioners described workflows that evolved from an offline into an online phase throughout development. 

\paragraph{\textbf{Offline Development}} First, practitioners coordinate with the legal and data teams to receive access to necessary data and begin the \textbf{data acquisition and processing} phase. The data team negotiates technical details such as API designs for data collection as well as data cleaning, while the legal team reviews data access requests. Practitioners then spend much of their time on \textbf{model construction} and conducting rounds of \textbf{model validation}. All participants agreed that they invested more than half of their work time into these model-centric activities. For example, P5 reported, ``\textit{This is the core of our daily work. I spend more than 60\% [of my] time on model work.}'' Overall, the offline stages of the RS practitioner workflow align with the pre-deployment stages of general ML pipelines as described by \citet{suresh_framework_2021} and \citet{black_toward_2023}, as well as with preferences towards ``model work'' over ``data work'' that have been previously documented by \citet{sambasivan_everyone_2021}.

\paragraph{\textbf{Online Development}} After offline development, practitioners ship their model and move to a \textbf{model testing} phase, where real-world user behavior data and newly generated user-content interaction data are collected as users engage with recommendations. A fairness team, which all participants reported exists in their organizations under various names (such as `responsible AI'), provides feedback to RS practitioners at this stage (see \S\ref{sec:fairness_workflow}). Simultaneously, practitioners monitor model performance metrics during the \textbf{model service} phase, and might revisit the data acquisition and pre-processing phase to change their training data (e.g., performing feature transformations or asking for more features). These online stages of the RS practitioner workflow align with the post-deployment stages of general ML pipelines \cite{suresh_framework_2021, black_toward_2023}. However, while prior work has identified only a small amount of ML fairness research focused on post-deployment efforts \cite{black_toward_2023}, participants described the online phases as a critical---and long-running---part of their workflow. As P2 put it, ``\textit{We are in [a] loop [where we keep] validating, online testing and improving the model.}''

\subsection{How the RS Practitioner Workflow Incorporates Fairness}\label{sec:fairness_workflow}

Fairness assessments and improvements occurred at several points throughout the RS workflow, often corresponding to shifts to new development stages. These transition points were convenient locations to incorporate fairness because they often included interactions between different teams. Within these transition points, the overall process for incorporating fairness considerations into the RS practitioner workflow takes shape. It consists of three major phases, as depicted in Figure~\ref{fig:workflow}: the prototyping phase, the internal phase, and the collaborative phase.

\paragraph{\textbf{Prototyping Phase}} The prototyping phase co-occurs with the data acquisition and processing phase of the RS workflow. In the data acquisition and processing phase, technical RS practitioners interact with legal and data teams to obtain access to training data. Here, procedural fairness considerations are raised by the first time as the legal team communicates to the technical  team which features (e.g., sensitive attributes) should not be used in building RS due to regulatory requirements or company policy. Unlike prior work, which has often found AI practitioners to be at odds with legal teams who see measuring unfairness as a potential legal liability \cite{costanza-chock_who_2022, wright_null_2024}, we found that participants had overwhelmingly positive opinions of the legal team (for example, P2 saw the legal team as ``\textit{help[ing] us comply well with our company's vision and values.}'').

Concurrently, practitioners spend time during the prototyping phase hypothesizing about the sources of unfairness that might exist in the training data and developing metrics intended to examine those hypotheses. Participants based hypotheses on their prior experiences with RS, and did not generally cite academic research as a source. P9 stated that even if they were aware of relevant academic research on fairness considerations related to their use case, they still ``\textit{need to come up with a clear formula and implement it by ourselves---it is really domain-specific.}'' Instead, participants reported reusing fairness metrics that they had previously developed to assess training data in this stage and emphasized that their initial metrics were ``\textit{intuition-based}'' (P3). A similar reliance on bespoke as opposed to standardized fairness metrics and approaches has been noted in prior work on NLP practitioners~\cite{harvey-etal-2025-understanding} as well as AI practitioners generally~\cite{madaio2022assessing, madaio_tinker_2024}.

\paragraph{\textbf{Internal Phase}} During the offline model validation phase of the RS workflow, practitioners moved to the internal phase of the fairness workflow, wherein they validated metrics developed during the prototyping phase and manually searched for unexpected results. Participants expressed that they undertake several rounds of iteration in this phase to find as many instances of potential unfairness as possible prior to model deployment. In addition, participants emphasized the uncertain nature of the internal phase. According to P1, ``\textit{It is really hard to predict users' behaviors---a lot of uncertainties.}'' P2 described the internal phase bluntly: ``\textit{We have to look carefully to find those hidden biases. It is not fun.}'' These comments point to a practical challenge that has not been widely discussed in prior work: because RS are an inherently collaborative technology, testing them requires a model of user behavior. Thus, relying on historical data as a test set, as is commonly done in prediction tasks, is not sufficient in an RS setting.

\paragraph{\textbf{Collaborative Phase}} After the model is shipped for online testing, the fairness workflow enters the collaborative phase. This phase is typically the first to include the fairness team, as well as the first phase to include data from real-world users of the RS. In the collaborative phase, the fairness team offers feedback based on their internal guidelines. Most participants reported that they had discretion as to whether to integrate feedback from the fairness team. Because participants felt that they had limited time and resources available, they sometimes attempted to weigh the relative impact of potential fairness issues against the work required to resolve them:
\begin{quote}
    \textit{P1: If [the fairness team's] feedback is about removing one particular feature, normally we will simply do it. However, if it requires a lot of work, we probably need to have another broader discussion.}
\end{quote}
As the owners of an RS within their organizations, participants reported numerous ongoing job requirements, such as monitoring online model performance. Participants primarily reported being concerned about factors such as uptime, and did not consider fairness to reach a similar level of disruptive emergency necessitating urgent correction. These types of competing demands on practitioners' time have been repeatedly noted by prior research~\cite{wang_strategies_2024, madaio2022assessing}. However, the iterative, manual, and user behavior-dependent aspects of fairness measurement are more specific to RS. When participants were able to address fairness issues in deployed models, they often had to integrate user feedback and data to seek out ``\textit{hidden}'' (P2, P7) types of unfairness (i.e., those that they had not anticipated in the prototyping stage). This process is highly iterative and requires much manual analysis. Participants reported that, in a typical project, they eventually composed over five new fairness metrics tailored to the specific system and context, even though they typically started with pre-existing metrics from prior projects. 

\section{Technical Challenges Arising Within RS Practitioner Teams When Incorporating Fairness} \label{sec:thematic}
\begin{table*}
\small
\centering
\begin{tabular}{>{\centering\hspace{0pt}}m{0.169\linewidth}>{\centering\hspace{0pt}}m{0.192\linewidth}>{\hspace{0pt}}m{0.579\linewidth}} 
\toprule
\textbf{Scope} & \textbf{Type of Challenge} & \multicolumn{1}{>{\centering\arraybackslash\hspace{0pt}}m{0.579\linewidth}}{\textbf{Challenge}} \\
\midrule

\multirow{3}{\linewidth}{\par{}\textbf{Technical}\newline\textit{(Within RS Teams)}} & 
Defining Fairness & 
Defining fairness in specific RS contexts, especially those not subject to existing rules or regulations\par{}\vspace{3pt}
Adapting pre-existing fairness metrics for RS, and managing the internal documentation involved in those adaptations\\
\cmidrule{2-3} & 
Considering Multiple Types of Fairness Across Many Groups & 
Managing (conflicting) multi-stakeholder interests\par{}\vspace{3pt}
Prioritizing between different (non-conflicting) types of fairness\par{}\vspace{3pt}
Handling high numbers of user groups\\
\cmidrule{2-3} & 
Dynamic Fairness Considerations & 
Developing ad hoc metrics to detect unexpected sources of bias\par{}\vspace{3pt}
Managing real-world user interaction and feedback loops \\
\midrule
\multirow{2}{\linewidth}{\textbf{Organizational }\\\textit{(Cross-Team)}} & 
Lack of Time & 
Finding time to work on fairness within technical teams and in cross-team collaborations \\
\cmidrule{2-3} & 
Cross-Team Communication & Communicating and collaborating effectively with external teams\\
\bottomrule
\end{tabular}
\caption{\textbf{Challenges Faced by RS Practitioners when Incorporating Fairness into their Workflows.} Challenges occur within RS teams as well as across teams and at the organization level.}\label{tab:challenges}
\Description{A table summarizing challenges faced by RS practitioners when incorporating fairness into their workflows. Technical challenges (within RS teams) include defining fairness, considering multiple types of fairness across many groups, and dynamic fairness considerations. Defining fairness includes defining fairness in specific RS contexts, especially those not subject to pre-existing rules or regulation, adapting pre-existing fairness metrics for RS, and managing the internal documentation involved in those adaptations. Considering multiple types of fairness across many groups includes managing (conflicting) multi-stakeholder interests, prioritizing between different (non-conflicting) types of fairness, and handling high numbers of user groups. Dynamic fairness considerations include developing ad hoc metrics to detect unexpected sources of bias, especially when managing real-world user interaction and feedback loops. Organizational challenges (cross-team) include lack of time (finding time to work on fairness within technical teams and in cross-team collaborations) and cross-team communication (communicating and collaborating effectively with external teams).
}
\end{table*}
In this section, we describe the key technical challenges faced by RS practitioners who seek to incorporate fairness considerations into their workflows (summarized in Table \ref{tab:challenges}). Overall, we find that defining fairness in the first place is a challenge, echoing findings from prior interview studies conducted on AI and ML practitioners~\cite{madaio2022assessing}. We also find challenges that are more unique to RS, such as that RS practitioners must consider multiple types of fairness across many different stakeholder groups. RS practitioners must not only balance conflicting stakeholder interests, but also prioritize between a large number of different \textit{non-conflicting} types of fairness. In addition, RS practitioners face unique challenges related to assessing and improving fairness in dynamic settings. Because RS are collaborative systems that are shaped by user behavior, pre-deployment fairness assessments are especially challenging.

\subsection{RS Practitioners Struggle to Define Fairness in the Context of Recommender Systems}\label{subsec:defining}
\paragraph{\textbf{Defining Fairness}} RS practitioners attempt to assess and improve fairness within two general categories. The first category is fairness as mandated by existing rules or regulations. For example, an RS that discriminates against job candidates of particular races would be considered unfair under US anti-discrimination law. Even when participants had a legally mandated fairness goal in mind, they reported struggling with how to achieve it. Participants felt they were not armed with specific tools or standardized methods to reach even well-specified fairness goals. This echoes \citet{smith2023scoping}, who found that as RS practitioners tried to select appropriate fairness metrics, they struggled with not only normative questions like which stakeholders should be prioritized, but also practical ones like whether to measure fairness across an entire distribution of recommendations or only across a top-k set of rankings.

Participants perceived the second category, encompassing all contexts not covered by existing rules or regulations, as more difficult and nuanced. This category includes cases where attributes that are legally protected in some contexts are not legally protected in others. For example, although racial discrimination is illegal in employment RS, it is not illegal in dating RS~\cite{hutson_debiasing_2018}. Thus, it is not clear how consistently to apply the same fairness metrics and techniques across contexts. This category also includes fairness across attributes that are not legally protected in any context, such as the popularity of a provider's content. These attributes frequently conflict and can compound to the point of being overwhelming. For example, P7 stated that, ``\textit{Each module of my recommender system requires different debiasing,}'' with goals ranging from avoiding a ``\textit{rich gets richer}'' feedback loop to providing recommendations of equal quality to different groups of end users. This question of what to do in the absence of legal requirements is one that industry AI teams consistently grapple with~\cite{quinonero_candela_disentangling_2023}.

\paragraph{\textbf{Adapting Pre-Existing Fairness Metrics and Navigating Documentation Debt}} While approaches for assessing and improving fairness in RS have been proposed by researchers (see \S\ref{subsec:background:fairML}), participants found them hard to directly adapt and ill-suited for their existing workflows. In some cases, participants simply did not have the time to find existing metrics:
\begin{quote}
    \textit{P4: I didn't find very useful literature but honestly, I didn't have much time to do a serious literature review and read papers either.}
\end{quote}
Participants also struggled to build and re-use institutional knowledge about fairness in RS. This was due to the fact that they faced documentation debt~\cite{bender_dangers_2021}: prior fairness-related work at their organizations was not typically well-documented for future use. Half of practitioners interviewed therefore faced difficulties in re-using internal metrics that had been previously developed for fairness purposes: 
\begin{quote}
    \textit{P7: For sure we can find [old] code and comments, but without some context archived, some documents, we are not sure whether we can use [them] directly.}
\end{quote}
Participants also felt ill-equipped to document their own fairness processes. As P3 stated, ``\textit{Documentation is a good idea\ldots however, I am not sure how specific [I should be].}'' These challenges strongly echo those surfaced by prior work on NLP, ML, and AI practitioners~\cite{harvey-etal-2025-understanding, holstein2019improving, ojewale_2025_towards, lee_landscape_2021}. 

\subsection{RS Practitioners Must Account for Multiple Types of Fairness Across Many Groups}
\paragraph{\textbf{Managing Conflicting Multi-Stakeholder Interests}} RS simultaneously connect and serve multiple stakeholders, including providers and end-users~\cite{burke2017multisided, li2021tutorial, smith2023scoping}. This multifaceted interaction introduces complex challenges, especially in the context of fairness. Participants commonly made tradeoffs, often prioritizing one user group over others. However, deciding how to make these tradeoffs was a challenge, as highlighted by P9, who felt that, ``\textit{Tradeoffs [between] groups of users\ldots are hard to [decide],}'' and P1, who pointed out a lack of a ``\textit{compelling [standard operating procedure] for those decisions.}'' Deciding which stakeholders to prioritize is a known challenge across fair ML \cite{madaio2022assessing}. However, in the context of RS, it may not always be necessary. Recent academic findings suggest that multi-objective optimization might not always adversely affect particular user groups, indicating a disconnect between industry practices and academic research in the RS domain~\cite{ekstrand2022fairness}.

\paragraph{\textbf{Prioritizing Between Different (Non-Conflicting) Types of Fairness}} Participants felt that there was more of a question of what to prioritize in the context of RS as compared to predictive models. In a predictive ML pipeline, fairness assessment typically focuses on one particular target (e.g., demographic parity). However, the practitioners that we interviewed expressed that they had to balance multiple fairness-related considerations. For example, one participant needed to consider demographic parity for job seekers in one region---and simultaneously ensure that advertisers across multiple regions were adhering to different sets of fairness-related regulatory requirements. Importantly, these fairness considerations were not necessarily in conflict with one another. However, because practitioners have limited time and resources available, they nevertheless need to choose certain fairness ideals to prioritize before turning to others:

\begin{quote}
    \textit{P3: Debiasing for multiple purposes is crucial in ensuring a fair and unbiased platform for all stakeholders. It's not just about the debiasing of content creators based on region, gender, and religion\ldots but also about addressing other factors that can impact the platform\ldots For instance, we need to make sure that products\ldots are supported fairly [e.g., that products that are artificially cheap because of government subsidies designed to gain market dominance are downranked on our RS platform].}
\end{quote} 
Because most work in fair ML focuses on prediction, in which different types of fairness are often incompatible with one another, prioritizing between different non-conflicting types of fairness represents a distinct challenge to doing fair ML in practice. While prior work has found that practitioners struggle to make value judgments about what to prioritize when faced with capacity constraints, that work tends to focus on the relationship between fairness and other, non-conflicting but not necessarily fairness-related, desiderata. For example, fairness efforts may be de-prioritized due to a focus on user privacy~\cite{holstein2019improving} or an organizational emphasis on maximizing user engagement or product functionality~\cite{rakova2021responsible}. Other work rightly notes that different fairness paradigms (e.g., procedural fairness and outcome-based fairness) do not necessarily conflict with one another and thus can be simultaneously satisfied by combining procedural fairness interventions (e.g., checklists) with technical ones intended to improve an outcome-based fairness metric~\cite{madaio2020co}. In contrast, we focus on the idea that even within a single paradigm (e.g., outcome-based fairness), there are multiple non-conflicting fairness definitions that RS systems could simultaneously satisfy. Nevertheless, we find that even when a given practitioner is tasked with improving fairness (i.e., when fairness in general is not de-prioritized), that practitioner must essentially rank their planned fairness interventions (including both procedural and outcome-focused interventions) and prioritize accordingly.

\paragraph{\textbf{Handling High Numbers of User Groups}} A significant challenge identified by participants involves ensuring fairness across a wide range of groups within RS platforms. Current metrics often fail to address the complexities arising from the large number of groups involved (and the ensuing intersectional groups). Multiple participants raised this challenge: 

\begin{quote}
    \textit{P2: It's hard to manage disparities between numerous groups. With thousands of groups to consider, it becomes nearly impossible to ensure uniformity across all.}
\end{quote}
   
\begin{quote}
    \textit{P7: Most of the training data is too sparse, and it is hard for us to enforce [fairness] between groups. I cannot look through thousands of groups to make sure [fairness] holds.}
\end{quote}
Participants also struggled to find metrics that went beyond group fairness. For example, P9 noted that ``\textit{Most metrics are designed with the assumption of a few groups\ldots They don't cater to intra-group-level debiasing, making our adaptation efforts cumbersome.}'' This challenge is exacerbated not only by the large scale of RS platforms, but also by the previously mentioned challenges of determining how to define and prioritize fairness metrics that are not required by law. In predictive ML settings, fairness assessments often take cues from anti-discrimination law by assessing differences in outcomes across legally protected demographic groups. In contrast, the RS practitioners we interviewed appear to rely less on anti-discrimination law as a blueprint for fairness assessments.

\subsection{Fairness Considerations are Complicated by Dynamic RS Environments}
\paragraph{\textbf{Developing Ad Hoc Metrics to Detect Unexpected Instances of Unfairness}} Practitioners engaged in building RS often create bespoke metrics in collaboration with their teams. Such approaches underscore the reliance on practitioner expertise and intuition in assessing and improving fairness, as highlighted by P6: ``\textit{In general, I would like to write hand-crafted code to validate my hypothesis---experience is key to debiasing.}'' Nevertheless, participants acknowledged that their initial metrics often fell short in detecting unfairness, and they therefore engaged in iterative refinement of those metrics based on user feedback:

\begin{quote}
   \textit{P1: It is not an easy job\ldots so we need to start from somewhere. After doing some debiasing based on [our] experience, we look forward to getting feedback from real-world users\ldots We do everything that we can do. However, it is not rare we detect some other bias after the model is [deployed].}
\end{quote}
   
\begin{quote}
   \textit{P7: It sounds a bit contradictory. We have a model first and keep debiasing the model based on our users' new data. Although [it] risks bringing in some confirmatory bias, it works for now.}
\end{quote}
Again, this highlights a challenge specific to measuring and improving fairness in RS: offline testing is inherently limited because it requires detailed predictions of how users will interact with deployed RS. This means that RS practitioners must either develop models of predicted user behavior or must wait until systems are deployed for comprehensive testing. This stands in stark contrast to most predictive ML systems, which are often testable using historical data, and even to generative AI systems, which may be tested through red teaming.

\paragraph{\textbf{Managing Real-World User Interactions and Feedback Loops}} Practitioners also face the challenge of assessing and improving fairness in deployed (as opposed to pre-deployment) RS. This challenge is particularly acute in RS given the volume of data, the subtleties of human interactions, and the potential for user feedback loops to amplify latent sources of unfairness. P1  emphasized the difficulty in uncovering unexpected fairness issues in deployed systems, stating, ``\textit{As the bias definitions are so different, and there are so many different user groups, those biases are hiding behind the systems that we built.}'' Despite finding academic metrics less directly applicable (see \S\ref{subsec:defining}),
participants acknowledged the value of research in providing a foundation for their decision-making processes. For example, P7 stated, ``\textit{Though sometimes [checklists] are not directly related, I can still check [them] and they help me think through my reasoning. It made me more confident in my decision.}'' At the same time, however, participants did not feel comfortable using pre-defined unfairness mitigation techniques (i.e., from academic research) on deployed systems. As P2 put it, ``\textit{There is unknown risk of using those approaches. If the model performance drops a lot, it is a disaster.}'' This challenge is exacerbated by the fact that there is a comparatively small amount of fair ML research focused on post-deployment fairness improvements~\cite{black_toward_2023}.

\section{Organizational Challenges Arising Across Teams When Incorporating Fairness}\label{sec:org}
In this section, we describe the key organizational challenges faced by RS practitioners who seek to incorporate fairness considerations into their workflows (summarized in Table \ref{tab:challenges}). The challenges we identify---namely, a lack of time to engage in fairness work and poor cross-team communication---echo those identified by prior interview studies conducted on AI and ML practitioners~\cite{harvey-etal-2025-understanding, deng2023investigating, rakova2021responsible, balayn__2023}. However, we also identify a notable exception to the challenge of cross-team communication: while the RS practitioners we interviewed struggled to communicate with fairness teams, they overwhelmingly described clear and productive communications with legal teams. Thus, in this section, we contrast between the two to offer new insight into why RS practitioners struggle to develop a ``fairness lingua franca.''

\subsection{RS Practitioners Lack Time to Engage in Collaborative Fairness Work}
Participants universally described their workflows as highly iterative and time-sensitive. As a result, most felt they had a very limited amount of time to focus on fairness, either internally or in collaborations with the fairness team: all reported spending 10\% or less of their time on fairness-related tasks. As P7 put it, ``\textit{The negative effect spurred from the platform I worked for before is not so `bad.' We don't have the incentive to invest half of our time\ldots to comprehensively debias.}'' When prompted to reflect on this allocation, most participants indicated that, while allocating more time for fairness-related tasks could be helpful, they were neutral about doing so:
\begin{quote}
    \textit{P1: Even though I wish I could have more time on this, I understand why this proportionally makes sense. From the perspective of our [organization], prioritizing efforts on improving `important' metrics brings more impact.}
\end{quote}
In other words, fairness issues did not rise to the level of a disruptive emergency, and thus were not prioritzed by management in evaluating project outcomes. This challenge is not unique to RS, and has been identified by researchers time and time again as a particular barrier to doing fairness work in practice~\cite{veale2018fairness, holstein2019improving, madaio2020co, rakova2021responsible, madaio2022assessing, deng2023investigating, balayn__2023, harvey-etal-2025-understanding}.

\subsection{RS Practitioners Lack a `Fairness Lingua Franca' to Facilitate Cross-Team Communication}

Engaging with external fairness teams revealed a nuanced challenge: attaining mutual understanding and converging perspectives. Participants valued fairness teams' feedback but sometimes struggled to integrate it due to a few key complexities. First, the complexity of RS sometimes made feedback from fairness teams difficult to implement, as highlighted by P1: \textit{``Their feedback could be something I agree with, but I cannot solve,''} and by P9: \textit{``Sometimes\ldots[a] challenge is fundamentally unsolvable, rooted in the deep learning models.''} Second, the multidisciplinary nature of fairness teams led them to use terminology that was not readily understood by technical teams, affecting inter-team communication:
\begin{quote}
    \textit{P6: Some of my colleagues are from the field of psychology and communication, and some of those terms I didn't know before.}
\end{quote}
\begin{quote}
    \textit{P9: Sometimes I don't understand phrases they use in their feedback, for example, `inductive bias from recommender systems,' and need to ask for more explanations.}
\end{quote}
Third, participants desired clearer guidance from fairness teams, particularly regarding model development. For example, P7 explained that, \textit{``Once [the fairness team] mentioned a certain aspect of my model might introduce bias, I wished to understand the reasoning.''} Finally, fairness teams were often involved only after models were deployed for online testing. As P9 put it, \textit{``I could work with [fairness] closer if they could have joined [our project] earlier.''}

We are not the first to identify communication challenges as a barrier to assessing and improving fairness in practice: notably, \citet{deng2023investigating} interviewed AI practitioners to understand ``cross-functional collaboration'' in industry. They found that practitioners addressed communication challenges by taking on ``bridging roles,'' in which they championed fairness efforts by initiating cross-team meetings and developing educational resources for their colleagues. While beneficial to their teams, practitioners who took on bridging roles struggled with under-recognition and unreasonable expectations, as their existing full-time jobs did not typically include explicit fairness-related responsibilities \cite{deng2023investigating}. Interestingly, even though the RS practitioners we interviewed all worked with dedicated fairness teams---theoretically reducing the need for informal bridging roles---they still struggled with cross-team communication.

Importantly, however, cross-team communication challenges were not ubiquitous: participants overwhelmingly drew a sharp contrast between their interactions with fairness teams and their interactions with legal teams. Due to the sensitive nature of end-user data, practitioners regularly interact with their legal team to confirm the scope of data that can be gathered. With increasing surveillance from authorities and global concerns over privacy in RS, one might expect the relationship between practitioners and internal policymakers to be fraught. However, participants reported that communication with legal teams was clear and direct. Per P2, ``\textit{If they said `no,' then it is a clear `no,' and we know the reasons and risks behind their decision.}'' Similarly, P3 stated, ``\textit{Their guidelines are very clear, particularly when compared with fairness guidelines, which we don't really have.}'' Clear and open guidelines from legal teams made it easier for practitioners to work with them, and most participants acknowledged the role of the legal team in avoiding risks. P1 summarized the issue succinctly: ``\textit{Our collaborations [with fairness teams] could be streamlined if there were guidelines as clear as those from legal teams.}'' Another thing that the technical practitioners we interviewed appreciated about legal teams was their early engagement---legal teams become involved in fairness work during the offline prototyping phase, while fairness teams typically do not engage until the online collaborative phase. Friction in cross-team understanding was a hindrance: 
\begin{quote}
    \textit{P9: Sometimes, [the fairness team asks] for a lot of abbreviations\ldots which can be found in our review documents. Personally, I could understand that joining in the middle of a project, it is hard for them to know a lot of discussions and terms that we used.}
\end{quote}
Similarly, P2 felt that fairness teams did not always provide actionable recommendations, saying ``\textit{[Fairness is] sometimes not on the same page---they gave feedback based on outdated [RS] designs.}'' These observations suggest potential paths forward for improving communication between fairness and technical teams.
\section{Calls to Action}\label{sec:calls}
In this section, we map the challenges uncovered in our interviews to calls for action, highlighting both where big tech companies and practitioners are responsible for implementing evidence-based changes to better integrate fairness into their RS workflows as well as where novel HCI research can support real-world fairness efforts.

\subsection{Building Institutional Knowledge Through Documentation}\label{subsect:documentation}
A key challenge reported by RS practitioners was adapting fairness metrics that had been previously developed at their organizations for new fairness work. Teams are composed of individuals from a wide variety of personal backgrounds, numerous project stages are spread over time, and systems are modified dynamically after deployment, when fairness effects may be hard to predict and can potentially self-reinforce in feedback loops. Due to these challenges, documenting fairness work is critical for building institutional fairness knowledge. However, as multiple practitioners pointed out, while there are standard practices for documenting and approving code changes, there is comparably little infrastructural support for documenting approaches to assess and improve fairness.

\paragraph{The Role of Practitioners and Organizations}
Practitioners and organizations must document their fairness metrics and prior fairness work. Practitioners and organizations are also responsible for effectively building on institutional knowledge. For example, the practitioners we interviewed reported that, when available, they re-used metrics from prior projects in the initial prototyping stage of the fairness workflow, but typically developed around five brand-new metrics by the end of the collaboration stage. This ad-hoc approach impedes the accumulation of institutional knowledge. It may harm procedural fairness as well, as different fairness standards are applied across different systems, and standards may not always be rigorously evaluated for validity. Thus, we suggest that practitioners should design metrics with extensibility, or the ability for metrics to be ``adapted for different systems, use cases, and deployment contexts'' in mind~\cite{harvey-etal-2025-understanding}.

\paragraph{The Role of HCI Researchers} HCI and ML fairness researchers have previously led calls for improved documentation in both research and industry. Researchers have proposed documentation frameworks including datasheets for datasets~\cite{gebru_datasheets_2021} and model cards for model reporting~\cite{mitchell_model_2019}---and have led efforts to understand the extent to which documentation meets the needs of and is adopted by industry practitioners~\cite{heger2022understanding, boyd2021datasheets}. More recently, researchers have begun to propose domain-specific augmentations to documentation frameworks, such as datasheets for speech datasets~\cite{papakyriakopoulos2023augmented} or for digital cultural heritage datasets~\cite{Alkemade2023}. Others have documented how AI practitioners themselves customize and adapt artifacts from research, like fairness checklists, to suit their needs and align with their workflows~\cite{madaio_tinker_2024, morley_operationalising_2023, madaio2020co}. Going forward, HCI researchers are uniquely well-suited to determining how such artifacts should be designed, i.e., which design practices lead to the development of adaptable or extensible documentation frameworks. Some researchers are already working on this: for example, \citet{madaio_tinker_2024} suggest that designers of artifacts intended for use by industry practitioners embrace ``positive ambiguity,'' or under-specification, so that artifacts can more easily be adapted. Similarly, \citet{harvey-etal-2025-understanding} propose that artifacts should be designed to be modifiable and modular in order to achieve extensibility. Both \citet{madaio_tinker_2024} and \citet{harvey-etal-2025-understanding} leave open the question of how ambiguous or modifiable documentation frameworks and other artifacts can be before they no longer perform their intended function of standardization, suggesting one clear avenue for future HCI research.

\subsection{Investing in Multi-Stakeholder Fairness}
Multi-stakeholder fairness considerations raised a variety of challenges, including managing conflicting stakeholder interests; prioritizing between many different, non-conflicting, types of fairness; and managing the technical and prioritization challenges associated with handling high numbers of user groups. 

\paragraph{The Role of Practitioners and Organizations} To tackle the challenges involved in improving fairness in RS, organizations must invest resources into researching, or working to implement academic research on, several key areas. First, it is necessary to develop more comprehensive optimization approaches, considering multiple objectives, to balance (or avoid) tradeoffs between different stakeholder groups. Second, research should address the ``too many groups'' challenge in RS by making it easier to assess fairness across a very high numbers of groups. Similarly, there should be an increased focus on intra-group fairness to align with multi-faceted and intersectional conceptions of diversity in the real world~\cite{wang_identities_2025, wang_towards_2022, Kanubala_Valera_2025}. While ML researchers can support this research, the massive scale of modern RS mean that much of this work will be computationally and financially expensive, and thus much of the responsibility should fall to well-resourced big tech organizations who also profit from RS~\cite{Widder2024}.

\paragraph{The Role of HCI Researchers} HCI researchers can advance multi-stakeholder fairness by developing transparent and standardized guidelines to scaffold decision-making when fairness tradeoffs between stakeholder groups are unavoidable. Value-sensitive design (VSD) methods, which embrace tensions between conflicting stakeholder values as a core theoretical construct, could prove a valuable approach~\cite{friedman2019value, boyd_designing_2022, zhu_value_2018, Simon2020-bf, shen_model_2022}. Both \citet{zhu_value_2018} and \citet{Simon2020-bf} have suggested that VSD's methodology of iterative conceptual, empirical, and technical investigations can be incorporated into the design of fair algorithms. As a concrete example, \citet{shen_model_2022} use a VSD approach to create the Model Card Authoring Toolkit, which is intended to help community members with ``tradeoff deliberation'' while designing decision-making algorithms. We suggest that HCI research focused on similar VSD-informed efforts could help avoid the ad hoc and potentially arbitrary decision-making that currently occurs when RS practitioners are forced to make fairness tradeoffs in real time. Guidelines should be developed in collaboration with diverse stakeholder groups, and should encourage practitioners to question, rather than reinforce, existing power dynamics on RS platforms.

\subsection{Integrating Fairness Considerations into Workflows Early and Often}
RS practitioners typically did not interact with fairness teams until models had been deployed. This delay caused friction, as fairness teams sometimes had out-of-date information on technical terminology or system updates. At the same time, during online phases of model development (when fairness teams were in the loop), technical teams were consistently ``\textit{on-call}'' (P2) and reported prioritizing maintaining model uptime over fairness fixes.

\paragraph{The Role of Practitioners and Organizations} On an organizational level, we found that practitioners encountered competing incentives for integrating fairness into their workflows. While practitioners' primary job responsibilities are to create and manage RS, every practitioner interviewed felt both a personal interest and an explicit requirement to assess and improve fairness in their models. At the same time, however, we noticed many cases where practitioners had to backtrack or iterate on deployed models in order to mitigate unexpected instances of unfairness. Had they invested additional time during project formation and received support from fairness teams earlier, perhaps such issues could have been detected when changes were easier to make. Our workflow model (Figures~\ref{fig:workflow}) highlights the opportunity for earlier engagement with fairness teams. The practitioners we interviewed explicitly stated that they would have preferred fairness teams to be involved with their projects earlier on; there is ample opportunity for this during the prototyping and internal phases of the fairness workflow.

\paragraph{The Role of HCI Researchers} Researchers are well-equipped to develop tools that support industry practitioners in incorporating fairness into their workflows. For example, \citet{Boenig-Liptsin02092022} have proposed the Ethos Lifecycle, an interactive online tool designed to increase the extent to which data science practitioners consider positionality, power, the interrelatedness of social and technical considerations, and the narratives animating their work. The tool is built around a model of a data science workflow and invites users to consider relevant ethical questions at each stage of the workflow. Similarly, \citet{boyd_designing_2022} uses VSD principles to create a field guide to inform practitioners about technical mitigations to ethical issues that they identify in the course of their work. That research was grounded in the concept of ethical sensitivity, which \citet{boyd_adapting_2021} define as ``the moment of noticing an ethical problem (recognition), the process of building understanding of the situation (particularization), and the decision about what to do (judgment).'' Following \citet{boyd_adapting_2021}, we suggest that additional HCI research is needed to understand the ways in which ethical recognition can be triggered and how such triggers can be built in to industry workflows. Furthermore, we suggest that HCI researchers can build on our model of the RS practitioner workflow to create teaching tools that could improve ethical particularization and judgment and encourage early consideration of fairness in the development of RS. 

\subsection{Developing a `Fairness Lingua Franca'}
We identified several cases where poor communication between fairness teams and technical practitioners led to confusion and even deprioritization of fairness efforts. 

\paragraph{The Role of Practitioners and Organizations} While legal teams offered explicit guidance and directives to practitioners, the nature of fairness feedback---reliant on nuance and not necessarily immediately applicable to complex models---proved much harder to parse and operationalize in systems. Combined with the fast-paced nature of RS development, this led to disconnects between technical practitioners and fairness teams despite mutual agreement on the importance of fairness. We believe that these disconnects could have been anticipated. Practitioners reported that fairness teams came from multiple disciplines, including some (like psychology and communication) which may not share common terminology with ML. RS practitioners lacked the broad interdisciplinary expertise represented on fairness teams, resulting in barriers to fully understanding the implications and context of fairness feedback. The dynamic environment of RS exacerbated this gap, leading practitioners to optimize by solving easy issues while neglecting other---potentially more serious---ones. Nevertheless, the gap could be reduced by more frequent meetings between technical and fairness teams in which shared terminologies are established. Likewise, embedding fairness practices more deeply in technical teams through additional roles, building domain expertise in mutual fields, and pursuing techniques for more consistent documentation and information flow (especially at the beginning of the RS workflow) might help to establish common ground.

We found that, when participants were aware of RS fairness research, research could provide basic principles for practitioners to follow or represent starting points for bespoke metrics. Yet several practitioners reported that they did not have the time to regularly review academic research, and thus that they consistently developed their own ad hoc fairness metrics instead. This creates a patchwork of practices that at times adheres to academic recommendations and at other times drifts far from them. While practitioners do not necessarily seek to disregard academic best practices, practical constraints sometimes make it hard to follow, or even know about, them. We suggest that the big tech companies that employ RS practitioners are responsible for incentivizing practitioners to devote the time necessary to stay up-to-date on academic research, i.e., by making it an explicit job responsibility as opposed to a form of invisible labor~\cite{deng2023investigating}.

\paragraph{The Role of HCI Researchers} We suggest that HCI researchers might help with the necessary translation work to provide common ground for technical and fairness practitioners. For example, \citet{Morley2021} propose Ethics as a Service, as a model for helping practitioners ``translate principles into practice.'' Ethics as a Service suggests concrete roles for organizations and practitioners to adopt organization-wide ethical principles and determine the appropriate approaches for implementing them. Similarly, \citet{delgado_uncommon_2022} identify factors that led to the successful translation of real-world legal requirements into an AI system through a participatory design task, including extended co-design and evaluation of the AI solutions and the creation of opportunities for expertise to ``flow in multiple directions.'' HCI researchers can build on this prior work to identify best practices for collaboration and information exchange during RS development to better support cross-team fairness efforts. A particularly useful starting point for this work is \citet{deng2023investigating}, who describe how practitioners take on ``bridging'' roles to build AI fairness expertise at their organizations and identify a variety of efforts undertaken by practitioners with fairness expertise to educate practitioners without that expertise, including hosting design workshops and hackathons and developing guidebooks and toolkits. At the same time, the work finds that bridging efforts are labor-intensive and under-appreciated~\cite{deng2023investigating}. Thus, we call for additional research into the effectiveness of different bridging strategies, as well as how bridging work can be made easier for practitioners.

\section{Discussion}\label{sec:discussion}

Despite the aforementioned challenges, we have reason to be optimistic: RS practitioners consistently ascribe importance to fairness work. In our semi-structured interviews, we asked practitioners to describe their emotional stress and satisfaction level in their daily work. While most practitioners reported that balancing daily work and fairness considerations was somewhat stressful, most also believed that fairness is an important and necessary part of their work. P9 described incorporating fairness as ``\textit{tiring but really rewarding.}'' Practitioners pointed out that fairness felt particularly important given the uniquely large impact that RS play on society:
\begin{quote}
    \textit{P1: It is tiring; however, I believe this is the right thing to do. It is important to remove bias as much as possible as our system serves millions of users worldwide. Discriminating behaviors imposed by our system might greatly affect their lives.}
\end{quote}
Practitioners expressed a desire to improve the world through their systems and identified strong intrinsic motivations for improving fairness, even if it is arduous. Some practitioners cited public and academic discourse on fairness as motivating:
\begin{quote}
    \textit{P1: Thorough discussions from existing papers, like COMPAS \cite{compas}, about how unfairness affects other people's life greatly shaped my view on this. As I have the chance to repair this, I will do my best---though it is not easy.}
\end{quote}

\subsection{Limitations and Ethical Considerations}\label{subsect:limitations}
\paragraph{Limitations} Our study is subject to a number of limitations. First and foremost, our findings and discussions are based on a limited sample of practitioners from seven highly visible companies. Although our sample size is typical of HCI interview studies~\cite{caine_local_2016}, and although we intentionally targeted practitioners working on very large-scale RS with high societal impact, this limits the kind of evidence that we were able to gather. In particular, our findings may not generalize to RS practitioners working on smaller-scale RS. Similarly, it is likely that fairness is more salient to the specific subset of individuals who agreed to be interviewed than it would be to a representative sample of RS practitioners. This does not necessarily imply that the individuals we interviewed feel differently about fairness than the average RS practitioner, but it is certainly possible that they were more likely to believe that fairness was ``\textit{rewarding}'' (P9) and ``\textit{the right thing to do}'' (P1) (see \S\ref{sec:discussion}). In any case, we do not attempt to draw generalizations about how common the fairness challenges and beliefs we identify are among RS practitioners as a whole---we only claim that these challenges exist, and that they impact the development of large-scale RS. Furthermore, this paper focuses on technical practitioners working on RS who are leading engineering or research efforts; we did not interview members of legal, data, or fairness teams. We chose to do this because we were specifically interested in how practitioners who are actively involved in building RS incorporate fairness into their responsibilities. Nevertheless, this does prevent us from analyzing the extent to which the workflows described by legal, data, and fairness teams align with---and deviate from---those described by technical teams. Further, our semi-structured interviews might be limited by the remote setting, potential risks of exposing corporate secrets, or participant self-presentation maintenance. While we found participants to generally be quite frank about their experiences, it is possible that they provided a more critical or a rosier picture of the environment due to their perceived expectations. Finally, the challenges we uncovered are likely not fully addressable: there is a fundamental tension between the desire for fairness rules that function similarly to rigid and straightforward legal and data governance requirements, and the ability to actually implement such rules in the wild~\cite{ali2023walking}. 

\paragraph{Ethical Considerations}
Our primary ethical consideration in conducting this study was related to the privacy and confidentiality of our interview subjects. In order to limit the possibility of re-identification, we provide only limited information about interview participants and their employers. We strictly adhered to our IRB guidelines during the interview and data analysis process by asking participants not to share sensitive information, de-identifying interview transcripts, and securely storing interview data.

\paragraph{Adverse Impacts}
While this study aims to foster understanding and improve fairness in RS, we acknowledge the potential for adverse impacts once published. The findings could unintentionally influence RS practitioners to adopt fairness practices that may not be universally applicable, leading to oversimplification of complex issues. Furthermore, there is a risk that the emphasis on certain strategies might overshadow other equally important aspects of fairness in RS. We recognize the importance of ongoing critical evaluation and discourse in the community to mitigate these potential negative impacts and encourage a diverse range of perspectives and solutions when addressing structural issues of fairness in RS.

\section{Conclusion}
In this work, we investigated fairness practices in RS through a semi-structured interview study of 11 technical practitioners who are building RS. We constructed a map of RS practitioners' daily workflow, focusing on how they interact with legal, data, and fairness teams at their organizations and on how they incorporate fairness into their existing practices. We identified key technical challenges to incorporating fairness, such as defining fairness in the context of RS and navigating the complexity introduced by RS; namely, multi-stakeholder fairness considerations and dynamic environments. We also identified organizational challenges resulting from cross-team interactions, including a lack of time to conduct fairness work and communication challenges across teams. We suggest that building institutional knowledge by improving documentation practices, encouraging investment into multi-stakeholder fairness approaches, integrating fairness considerations into RS workflows early and often, and developing a `fairness lingua franca' can lead to improvements in RS fairness.

\begin{acks}
This work was supported by NSF grant IIS-1850195. We would like to thank the associate chairs and anonymous reviewers for their invaluable feedback.  We also would like to thank the Dataprep team from Simon Fraser University for the valuable input of mapping common functions and user intentions. All opinions, findings, and conclusions in this paper are those of the authors and do not necessarily reflect the views of the funding agencies.
\end{acks}

\bibliographystyle{ACM-Reference-Format}
\bibliography{references}

@inproceedings{boyd_designing_2022,
author = {Boyd, Karen},
title = {Designing Up with Value-Sensitive Design: Building a Field Guide for Ethical ML Development},
year = {2022},
isbn = {9781450393522},
publisher = {Association for Computing Machinery},
address = {New York, NY, USA},
url = {https://doi.org/10.1145/3531146.3534626},
doi = {10.1145/3531146.3534626},
booktitle = {Proceedings of the 2022 ACM Conference on Fairness, Accountability, and Transparency},
pages = {2069–2082},
numpages = {14},
location = {Seoul, Republic of Korea},
series = {FAccT '22}
}

@inproceedings{shen_model_2022,
author = {Shen, Hong and Wang, Leijie and Deng, Wesley H. and Brusse, Ciell and Velgersdijk, Ronald and Zhu, Haiyi},
title = {The Model Card Authoring Toolkit: Toward Community-centered, Deliberation-driven AI Design},
year = {2022},
isbn = {9781450393522},
publisher = {Association for Computing Machinery},
address = {New York, NY, USA},
url = {https://doi.org/10.1145/3531146.3533110},
doi = {10.1145/3531146.3533110},
booktitle = {Proceedings of the 2022 ACM Conference on Fairness, Accountability, and Transparency},
pages = {440–451},
numpages = {12},
location = {Seoul, Republic of Korea},
series = {FAccT '22}
}

@ARTICLE{Simon2020-bf,
  title     = "Algorithmic bias and the Value Sensitive Design approach",
  author    = "Simon, Judith and Wong, Pak-Hang and Rieder, Gernot",
  journal   = "Internet Pol. Rev.",
  publisher = "Internet Policy Review, Alexander von Humboldt Institute for
               Internet and Society",
  volume    =  9,
  number    =  4,
  month     =  dec,
  year      =  2020,
  language  = "en"
}

@article{zhu_value_2018,
author = {Zhu, Haiyi and Yu, Bowen and Halfaker, Aaron and Terveen, Loren},
title = {Value-Sensitive Algorithm Design: Method, Case Study, and Lessons},
year = {2018},
issue_date = {November 2018},
publisher = {Association for Computing Machinery},
address = {New York, NY, USA},
volume = {2},
number = {CSCW},
url = {https://doi.org/10.1145/3274463},
doi = {10.1145/3274463},
journal = {Proc. ACM Hum.-Comput. Interact.},
month = nov,
articleno = {194},
numpages = {23}
}

@article{lee_webuildai_2019,
	series = {{CSCW}},
	title = {{WeBuildAI}: {Participatory} {Framework} for {Algorithmic} {Governance}},
	volume = {3},
	issn = {2573-0142},
	shorttitle = {{WeBuildAI}},
	url = {https://dl.acm.org/doi/10.1145/3359283},
	doi = {10.1145/3359283},
	language = {en},
	number = {CSCW},
	urldate = {2024-01-25},
	journal = {Proceedings of the ACM on Human-Computer Interaction},
	author = {Lee, Min Kyung and Kusbit, Daniel and Kahng, Anson and Kim, Ji Tae and Yuan, Xinran and Chan, Allissa and See, Daniel and Noothigattu, Ritesh and Lee, Siheon and Psomas, Alexandros and Procaccia, Ariel D.},
	month = nov,
	year = {2019},
	pages = {1--35},
}

@inproceedings{kunkel_let_2019,
author = {Kunkel, Johannes and Donkers, Tim and Michael, Lisa and Barbu, Catalin-Mihai and Ziegler, J\"{u}rgen},
title = {Let Me Explain: Impact of Personal and Impersonal Explanations on Trust in Recommender Systems},
year = {2019},
isbn = {9781450359702},
publisher = {Association for Computing Machinery},
address = {New York, NY, USA},
url = {https://doi-org.proxy.library.cornell.edu/10.1145/3290605.3300717},
doi = {10.1145/3290605.3300717},
booktitle = {Proceedings of the 2019 CHI Conference on Human Factors in Computing Systems},
pages = {1–12},
numpages = {12},
location = {Glasgow, Scotland Uk},
series = {CHI '19}
}

@article{gebru_datasheets_2021,
	title = {Datasheets for datasets},
	volume = {64},
	issn = {0001-0782, 1557-7317},
	url = {https://dl.acm.org/doi/10.1145/3458723},
	doi = {10.1145/3458723},
	language = {en},
	number = {12},
	urldate = {2023-04-07},
	journal = {Communications of the ACM},
	author = {Gebru, Timnit and Morgenstern, Jamie and Vecchione, Briana and Vaughan, Jennifer Wortman and Wallach, Hanna and Iii, Hal Daumé and Crawford, Kate},
	month = dec,
	year = {2021},
	pages = {86--92},
}

@inproceedings{Biega_equity_2018,
author = {Biega, Asia J. and Gummadi, Krishna P. and Weikum, Gerhard},
title = {Equity of Attention: Amortizing Individual Fairness in Rankings},
year = {2018},
isbn = {9781450356572},
publisher = {Association for Computing Machinery},
address = {New York, NY, USA},
url = {https://doi.org/10.1145/3209978.3210063},
doi = {10.1145/3209978.3210063},
booktitle = {The 41st International ACM SIGIR Conference on Research \& Development in Information Retrieval},
pages = {405–414},
numpages = {10},
location = {Ann Arbor, MI, USA},
series = {SIGIR '18}
}

@InProceedings{pmlr-v81-ekstrand18b,
  title = 	 {All The Cool Kids, How Do They Fit In?: Popularity and Demographic Biases in Recommender Evaluation and Effectiveness},
  author = 	 {Ekstrand, Michael D. and Tian, Mucun and Azpiazu, Ion Madrazo and Ekstrand, Jennifer D. and Anuyah, Oghenemaro and McNeill, David and Pera, Maria Soledad},
  booktitle = 	 {Proceedings of the 1st Conference on Fairness, Accountability and Transparency},
  pages = 	 {172--186},
  year = 	 {2018},
  editor = 	 {Friedler, Sorelle A. and Wilson, Christo},
  volume = 	 {81},
  series = 	 {Proceedings of Machine Learning Research},
  month = 	 {23--24 Feb},
  publisher =    {PMLR},
  url = 	 {https://proceedings.mlr.press/v81/ekstrand18b.html},
}

@inproceedings{li_user_2021,
author = {Li, Yunqi and Chen, Hanxiong and Fu, Zuohui and Ge, Yingqiang and Zhang, Yongfeng},
title = {User-oriented Fairness in Recommendation},
year = {2021},
isbn = {9781450383127},
publisher = {Association for Computing Machinery},
address = {New York, NY, USA},
url = {https://doi.org/10.1145/3442381.3449866},
doi = {10.1145/3442381.3449866},
booktitle = {Proceedings of the Web Conference 2021},
pages = {624–632},
numpages = {9},
location = {Ljubljana, Slovenia},
series = {WWW '21}
}

@article{zehlike_fairness_2022a,
author = {Zehlike, Meike and Yang, Ke and Stoyanovich, Julia},
title = {Fairness in Ranking, Part I: Score-Based Ranking},
year = {2022},
issue_date = {June 2023},
publisher = {Association for Computing Machinery},
address = {New York, NY, USA},
volume = {55},
number = {6},
issn = {0360-0300},
url = {https://doi.org/10.1145/3533379},
doi = {10.1145/3533379},
journal = {ACM Comput. Surv.},
month = dec,
articleno = {118},
numpages = {36}
}

@article{zehlike_fairness_2022b,
author = {Zehlike, Meike and Yang, Ke and Stoyanovich, Julia},
title = {Fairness in Ranking, Part II: Learning-to-Rank and Recommender Systems},
year = {2022},
issue_date = {June 2023},
publisher = {Association for Computing Machinery},
address = {New York, NY, USA},
volume = {55},
number = {6},
issn = {0360-0300},
url = {https://doi.org/10.1145/3533380},
doi = {10.1145/3533380},
journal = {ACM Comput. Surv.},
month = dec,
articleno = {117},
numpages = {41}
}

@inproceedings{rastogi_fairness_2024,
author = {Rastogi, Richa and Joachims, Thorsten},
title = {Fairness in Ranking under Disparate Uncertainty},
year = {2024},
isbn = {9798400712227},
publisher = {Association for Computing Machinery},
address = {New York, NY, USA},
url = {https://doi.org/10.1145/3689904.3694703},
doi = {10.1145/3689904.3694703},
booktitle = {Proceedings of the 4th ACM Conference on Equity and Access in Algorithms, Mechanisms, and Optimization},
articleno = {5},
numpages = {31},
location = {San Luis Potosi, Mexico},
series = {EAAMO '24}
}

@inproceedings{ye_measuring_2017,
author = {Yang, Ke and Stoyanovich, Julia},
title = {Measuring Fairness in Ranked Outputs},
year = {2017},
isbn = {9781450352826},
publisher = {Association for Computing Machinery},
address = {New York, NY, USA},
url = {https://doi.org/10.1145/3085504.3085526},
doi = {10.1145/3085504.3085526},
booktitle = {Proceedings of the 29th International Conference on Scientific and Statistical Database Management},
articleno = {22},
numpages = {6},
location = {Chicago, IL, USA},
series = {SSDBM '17}
}

@inproceedings{steck_calibrated_2018,
author = {Steck, Harald},
title = {Calibrated recommendations},
year = {2018},
isbn = {9781450359016},
publisher = {Association for Computing Machinery},
address = {New York, NY, USA},
url = {https://doi.org/10.1145/3240323.3240372},
doi = {10.1145/3240323.3240372},
booktitle = {Proceedings of the 12th ACM Conference on Recommender Systems},
pages = {154–162},
numpages = {9},
location = {Vancouver, British Columbia, Canada},
series = {RecSys '18}
}

@inproceedings{singh_fairness_2018,
author = {Singh, Ashudeep and Joachims, Thorsten},
title = {Fairness of Exposure in Rankings},
year = {2018},
isbn = {9781450355520},
publisher = {Association for Computing Machinery},
address = {New York, NY, USA},
url = {https://doi.org/10.1145/3219819.3220088},
doi = {10.1145/3219819.3220088},
booktitle = {Proceedings of the 24th ACM SIGKDD International Conference on Knowledge Discovery \& Data Mining},
pages = {2219–2228},
numpages = {10},
location = {London, United Kingdom},
series = {KDD '18}
}

@inproceedings{zehlike_reducing_2020,
author = {Zehlike, Meike and Castillo, Carlos},
title = {Reducing Disparate Exposure in Ranking: A Learning To Rank Approach},
year = {2020},
isbn = {9781450370233},
publisher = {Association for Computing Machinery},
address = {New York, NY, USA},
url = {https://doi.org/10.1145/3366424.3380048},
doi = {10.1145/3366424.3380048},
booktitle = {Proceedings of The Web Conference 2020},
pages = {2849–2855},
numpages = {7},
location = {Taipei, Taiwan},
series = {WWW '20}
}

@inproceedings{geyik_fairness_2019,
author = {Geyik, Sahin Cem and Ambler, Stuart and Kenthapadi, Krishnaram},
title = {Fairness-Aware Ranking in Search \& Recommendation Systems with Application to LinkedIn Talent Search},
year = {2019},
isbn = {9781450362016},
publisher = {Association for Computing Machinery},
address = {New York, NY, USA},
url = {https://doi.org/10.1145/3292500.3330691},
doi = {10.1145/3292500.3330691},
booktitle = {Proceedings of the 25th ACM SIGKDD International Conference on Knowledge Discovery \& Data Mining},
pages = {2221–2231},
numpages = {11},
location = {Anchorage, AK, USA},
series = {KDD '19}
}

@inproceedings{diaz_evaluating_2020,
author = {Diaz, Fernando and Mitra, Bhaskar and Ekstrand, Michael D. and Biega, Asia J. and Carterette, Ben},
title = {Evaluating Stochastic Rankings with Expected Exposure},
year = {2020},
isbn = {9781450368599},
publisher = {Association for Computing Machinery},
address = {New York, NY, USA},
url = {https://doi.org/10.1145/3340531.3411962},
doi = {10.1145/3340531.3411962},
booktitle = {Proceedings of the 29th ACM International Conference on Information \& Knowledge Management},
pages = {275–284},
numpages = {10},
location = {Virtual Event, Ireland},
series = {CIKM '20}
}

@inproceedings{beutel_fairness_2019,
author = {Beutel, Alex and Chen, Jilin and Doshi, Tulsee and Qian, Hai and Wei, Li and Wu, Yi and Heldt, Lukasz and Zhao, Zhe and Hong, Lichan and Chi, Ed H. and Goodrow, Cristos},
title = {Fairness in Recommendation Ranking through Pairwise Comparisons},
year = {2019},
isbn = {9781450362016},
publisher = {Association for Computing Machinery},
address = {New York, NY, USA},
url = {https://doi.org/10.1145/3292500.3330745},
doi = {10.1145/3292500.3330745},
booktitle = {Proceedings of the 25th ACM SIGKDD International Conference on Knowledge Discovery \& Data Mining},
pages = {2212–2220},
numpages = {9},
location = {Anchorage, AK, USA},
series = {KDD '19}
}

@article{ali_discrimination_2019,
author = {Ali, Muhammad and Sapiezynski, Piotr and Bogen, Miranda and Korolova, Aleksandra and Mislove, Alan and Rieke, Aaron},
title = {Discrimination through Optimization: How Facebook's Ad Delivery Can Lead to Biased Outcomes},
year = {2019},
issue_date = {November 2019},
publisher = {Association for Computing Machinery},
address = {New York, NY, USA},
volume = {3},
number = {CSCW},
url = {https://doi.org/10.1145/3359301},
doi = {10.1145/3359301},
journal = {Proc. ACM Hum.-Comput. Interact.},
month = nov,
articleno = {199},
numpages = {30}
}

@inproceedings{asudeh_designing_2019,
author = {Asudeh, Abolfazl and Jagadish, H. V. and Stoyanovich, Julia and Das, Gautam},
title = {Designing Fair Ranking Schemes},
year = {2019},
isbn = {9781450356435},
publisher = {Association for Computing Machinery},
address = {New York, NY, USA},
url = {https://doi.org/10.1145/3299869.3300079},
doi = {10.1145/3299869.3300079},
booktitle = {Proceedings of the 2019 International Conference on Management of Data},
pages = {1259–1276},
numpages = {18},
location = {Amsterdam, Netherlands},
series = {SIGMOD '19}
}

@inproceedings{Zehlike_fair_2017,
author = {Zehlike, Meike and Bonchi, Francesco and Castillo, Carlos and Hajian, Sara and Megahed, Mohamed and Baeza-Yates, Ricardo},
title = {FA*IR: A Fair Top-k Ranking Algorithm},
year = {2017},
isbn = {9781450349185},
publisher = {Association for Computing Machinery},
address = {New York, NY, USA},
url = {https://doi.org/10.1145/3132847.3132938},
doi = {10.1145/3132847.3132938},
booktitle = {Proceedings of the 2017 ACM on Conference on Information and Knowledge Management},
pages = {1569–1578},
numpages = {10},
location = {Singapore, Singapore},
series = {CIKM '17}
}

@inproceedings{patro_fair_2022,
author = {Patro, Gourab K. and Porcaro, Lorenzo and Mitchell, Laura and Zhang, Qiuyue and Zehlike, Meike and Garg, Nikhil},
title = {Fair ranking: a critical review, challenges, and future directions},
year = {2022},
isbn = {9781450393522},
publisher = {Association for Computing Machinery},
address = {New York, NY, USA},
url = {https://doi.org/10.1145/3531146.3533238},
doi = {10.1145/3531146.3533238},
booktitle = {Proceedings of the 2022 ACM Conference on Fairness, Accountability, and Transparency},
pages = {1929–1942},
numpages = {14},
location = {Seoul, Republic of Korea},
series = {FAccT '22}
}

@Inbook{Magrani2024,
author="Magrani, Eduardo
and da Silva, Paula Guedes Fernandes",
editor="Sousa Antunes, Henrique
and Freitas, Pedro Miguel
and Oliveira, Arlindo L.
and Martins Pereira, Clara
and Vaz de Sequeira, Elsa
and Barreto Xavier, Lu{\'i}s",
title="The Ethical and Legal Challenges of Recommender Systems Driven by Artificial Intelligence",
bookTitle="Multidisciplinary Perspectives on Artificial Intelligence and the Law",
year="2024",
publisher="Springer International Publishing",
address="Cham",
pages="141--168",
isbn="978-3-031-41264-6",
doi="10.1007/978-3-031-41264-6_8",
url="https://doi.org/10.1007/978-3-031-41264-6_8"
}

@ARTICLE{hildebrandt_issue_2022,
  
AUTHOR={Hildebrandt, Mireille },
         
TITLE={The Issue of Proxies and Choice Architectures. Why EU Law Matters for Recommender Systems},
        
JOURNAL={Frontiers in Artificial Intelligence},
        
VOLUME={Volume 5 - 2022},

YEAR={2022},

URL={https://www.frontiersin.org/journals/artificial-intelligence/articles/10.3389/frai.2022.789076},

DOI={10.3389/frai.2022.789076},

ISSN={2624-8212}}

@article{wong_seeing_2023,
	series = {{CSCW}},
	title = {Seeing {Like} a {Toolkit}: {How} {Toolkits} {Envision} the {Work} of {AI} {Ethics}},
	volume = {7},
	issn = {2573-0142},
	shorttitle = {Seeing {Like} a {Toolkit}},
	url = {https://dl.acm.org/doi/10.1145/3579621},
	doi = {10.1145/3579621},
	language = {en},
	number = {CSCW1},
	urldate = {2024-03-06},
	journal = {Proceedings of the ACM on Human-Computer Interaction},
	author = {Wong, Richmond Y. and Madaio, Michael A. and Merrill, Nick},
	month = apr,
	year = {2023},
	pages = {1--27},
}

@Article{Morley2020,
author={Morley, Jessica
and Floridi, Luciano
and Kinsey, Libby
and Elhalal, Anat},
title={From What to How: An Initial Review of Publicly Available AI Ethics Tools, Methods and Research to Translate Principles into Practices},
journal={Science and Engineering Ethics},
year={2020},
month={Aug},
day={01},
volume={26},
number={4},
pages={2141-2168},
issn={1471-5546},
doi={10.1007/s11948-019-00165-5},
url={https://doi.org/10.1007/s11948-019-00165-5}
}

@misc{compas, 
title={Machine Bias},
author={Julia Angwin and Jeff Larson and Surya Mattu and Lauren Kirchner},
year={2016},
url={https://www.propublica.org/article/machine-bias-risk-assessments-in-criminal-sentencing},
howpublished={ProPublica}}

@article{delgado_uncommon_2022,
	series = {{CSCW}},
	title = {An {Uncommon} {Task}: {Participatory} {Design} in {Legal} {AI}},
	volume = {6},
	issn = {2573-0142},
	shorttitle = {An {Uncommon} {Task}},
	url = {https://dl.acm.org/doi/10.1145/3512898},
	doi = {10.1145/3512898},
	language = {en},
	number = {CSCW1},
	urldate = {2023-07-07},
	journal = {Proceedings of the ACM on Human-Computer Interaction},
	author = {Delgado, Fernando and Barocas, Solon and Levy, Karen},
	month = mar,
	year = {2022},
	pages = {1--23},
}

@Article{Morley2021,
author={Morley, Jessica
and Elhalal, Anat
and Garcia, Francesca
and Kinsey, Libby
and M{\"o}kander, Jakob
and Floridi, Luciano},
title={Ethics as a Service: A Pragmatic Operationalisation of AI Ethics},
journal={Minds and Machines},
year={2021},
month={Jun},
day={01},
volume={31},
number={2},
pages={239-256},
issn={1572-8641},
doi={10.1007/s11023-021-09563-w},
url={https://doi.org/10.1007/s11023-021-09563-w}
}

@article{madaio_tinker_2024,
author = {Madaio, Michael A. and Chen, Jingya and Wallach, Hanna and Wortman Vaughan, Jennifer},
title = {Tinker, Tailor, Configure, Customize: The Articulation Work of Contextualizing an AI Fairness Checklist},
year = {2024},
issue_date = {April 2024},
publisher = {Association for Computing Machinery},
address = {New York, NY, USA},
volume = {8},
number = {CSCW1},
url = {https://doi.org/10.1145/3653705},
doi = {10.1145/3653705},
journal = {Proc. ACM Hum.-Comput. Interact.},
month = apr,
articleno = {214},
numpages = {20}
}

@inproceedings{mitchell_model_2019,
	address = {New York, NY, USA},
	series = {{FAT}*},
	title = {Model {Cards} for {Model} {Reporting}},
	isbn = {978-1-4503-6125-5},
	url = {https://doi.org/10.1145/3287560.3287596},
	doi = {10.1145/3287560.3287596},
	urldate = {2023-04-11},
	booktitle = {Proceedings of the {Conference} on {Fairness}, {Accountability}, and {Transparency}},
	publisher = {Association for Computing Machinery},
	author = {Mitchell, Margaret and Wu, Simone and Zaldivar, Andrew and Barnes, Parker and Vasserman, Lucy and Hutchinson, Ben and Spitzer, Elena and Raji, Inioluwa Deborah and Gebru, Timnit},
	month = jan,
	year = {2019},
	pages = {220--229},
}

@misc{arnold2019factsheetsincreasingtrustai,
      title={FactSheets: Increasing Trust in AI Services through Supplier's Declarations of Conformity}, 
      author={Matthew Arnold and Rachel K. E. Bellamy and Michael Hind and Stephanie Houde and Sameep Mehta and Aleksandra Mojsilovic and Ravi Nair and Karthikeyan Natesan Ramamurthy and Darrell Reimer and Alexandra Olteanu and David Piorkowski and Jason Tsay and Kush R. Varshney},
      year={2019},
      eprint={1808.07261},
      archivePrefix={arXiv},
      primaryClass={cs.CY},
      url={https://arxiv.org/abs/1808.07261}, 
}

@inproceedings{sun_when_2023,
author = {Sun, Yuan and Drivas, Magdalayna and Liao, Mengqi and Sundar, S. Shyam},
title = {When Recommender Systems Snoop into Social Media, Users Trust them Less for Health Advice},
year = {2023},
isbn = {9781450394215},
publisher = {Association for Computing Machinery},
address = {New York, NY, USA},
url = {https://doi.org/10.1145/3544548.3581123},
doi = {10.1145/3544548.3581123},
booktitle = {Proceedings of the 2023 CHI Conference on Human Factors in Computing Systems},
articleno = {818},
numpages = {14},
location = {Hamburg, Germany},
series = {CHI '23}
}

@inproceedings{jin_beyond_2024,
	address = {Honolulu HI USA},
	series = {{CHI}},
	title = {({Beyond}) {Reasonable} {Doubt}: {Challenges} that {Public} {Defenders} {Face} in {Scrutinizing} {AI} in {Court}},
	isbn = {979-8-4007-0330-0},
	shorttitle = {({Beyond}) {Reasonable} {Doubt}},
	url = {https://dl.acm.org/doi/10.1145/3613904.3641902},
	doi = {10.1145/3613904.3641902},
	language = {en},
	urldate = {2024-05-16},
	booktitle = {Proceedings of the {CHI} {Conference} on {Human} {Factors} in {Computing} {Systems}},
	publisher = {ACM},
	author = {Jin, Angela and Salehi, Niloufar},
	month = may,
	year = {2024},
	pages = {1--19},
}

@article{hutson_debiasing_2018,
author = {Hutson, Jevan A. and Taft, Jessie G. and Barocas, Solon and Levy, Karen},
title = {Debiasing Desire: Addressing Bias \& Discrimination on Intimate Platforms},
year = {2018},
issue_date = {November 2018},
publisher = {Association for Computing Machinery},
address = {New York, NY, USA},
volume = {2},
number = {CSCW},
url = {https://doi.org/10.1145/3274342},
doi = {10.1145/3274342},
journal = {Proc. ACM Hum.-Comput. Interact.},
month = nov,
articleno = {73},
numpages = {18}
}

@inproceedings{blodgett_language_2020,
	address = {Online},
	series = {{ACL}},
	title = {Language ({Technology}) is {Power}: {A} {Critical} {Survey} of “{Bias}” in {NLP}},
	shorttitle = {Language ({Technology}) is {Power}},
	url = {https://www.aclweb.org/anthology/2020.acl-main.485},
	doi = {10.18653/v1/2020.acl-main.485},
	language = {en},
	urldate = {2024-03-07},
	booktitle = {Proceedings of the 58th {Annual} {Meeting} of the {Association} for {Computational} {Linguistics}},
	publisher = {Association for Computational Linguistics},
	author = {Blodgett, Su Lin and Barocas, Solon and Daumé Iii, Hal and Wallach, Hanna},
	month = jul,
	year = {2020},
	pages = {5454--5476},
}

@inproceedings{lustig_designing_2022,
author = {Lustig, Caitlin and Konrad, Artie and Brubaker, Jed R.},
title = {Designing for the Bittersweet: Improving Sensitive Experiences with Recommender Systems},
year = {2022},
isbn = {9781450391573},
publisher = {Association for Computing Machinery},
address = {New York, NY, USA},
url = {https://doi.org/10.1145/3491102.3502049},
doi = {10.1145/3491102.3502049},
booktitle = {Proceedings of the 2022 CHI Conference on Human Factors in Computing Systems},
articleno = {16},
numpages = {18},
location = {New Orleans, LA, USA},
series = {CHI '22}
}

@article{wang_biased_2023,
author = {Wang, Clarice and Wang, Kathryn and Bian, Andrew Y. and Islam, Rashidul and Keya, Kamrun Naher and Foulds, James and Pan, Shimei},
title = {When Biased Humans Meet Debiased AI: A Case Study in College Major Recommendation},
year = {2023},
issue_date = {September 2023},
publisher = {Association for Computing Machinery},
address = {New York, NY, USA},
volume = {13},
number = {3},
issn = {2160-6455},
url = {https://doi.org/10.1145/3611313},
doi = {10.1145/3611313},
journal = {ACM Trans. Interact. Intell. Syst.},
month = sep,
articleno = {17},
numpages = {28}
}

@inproceedings{krause_effect_2025,
author = {Krause, Thorsten and G\"{o}ritz, Lorena and Gratz, Robin},
title = {The Effect of Gender De-biased Recommendations — A User Study on Gender-specific Preferences},
year = {2025},
isbn = {9798400713941},
publisher = {Association for Computing Machinery},
address = {New York, NY, USA},
url = {https://doi.org/10.1145/3706598.3713155},
doi = {10.1145/3706598.3713155},
booktitle = {Proceedings of the 2025 CHI Conference on Human Factors in Computing Systems},
articleno = {1000},
numpages = {16},
location = {
},
series = {CHI '25}
}

@book{rawls_theory_1999,
	address = {Cambridge, Mass},
	edition = {Rev. ed},
	title = {A theory of justice},
	isbn = {978-0-674-00077-3 978-0-674-00078-0},
	publisher = {Belknap Press of Harvard University Press},
	author = {Rawls, John},
	year = {1999},
}

@inproceedings{jacobs_measurement_2021,
	address = {New York, NY, USA},
	series = {{FAccT}},
	title = {Measurement and {Fairness}},
	isbn = {978-1-4503-8309-7},
	url = {https://dl.acm.org/doi/10.1145/3442188.3445901},
	doi = {10.1145/3442188.3445901},
	urldate = {2023-04-06},
	booktitle = {Proceedings of the 2021 {ACM} {Conference} on {Fairness}, {Accountability}, and {Transparency}},
	publisher = {Association for Computing Machinery},
	author = {Jacobs, Abigail Z. and Wallach, Hanna},
	month = mar,
	year = {2021},
	pages = {375--385},
}

@inproceedings{anthis-etal-2025-impossibility,
    title = "The Impossibility of Fair {LLM}s",
    author = "Anthis, Jacy Reese  and
      Lum, Kristian  and
      Ekstrand, Michael  and
      Feller, Avi  and
      Tan, Chenhao",
    editor = "Che, Wanxiang  and
      Nabende, Joyce  and
      Shutova, Ekaterina  and
      Pilehvar, Mohammad Taher",
    booktitle = "Proceedings of the 63rd Annual Meeting of the Association for Computational Linguistics (Volume 1: Long Papers)",
    month = jul,
    year = "2025",
    address = "Vienna, Austria",
    publisher = "Association for Computational Linguistics",
    url = "https://aclanthology.org/2025.acl-long.5/",
    doi = "10.18653/v1/2025.acl-long.5",
    pages = "105--120",
    ISBN = "979-8-89176-251-0",
}

@inproceedings{eslami_first_2016,
author = {Eslami, Motahhare and Karahalios, Karrie and Sandvig, Christian and Vaccaro, Kristen and Rickman, Aimee and Hamilton, Kevin and Kirlik, Alex},
title = {First I "like" it, then I hide it: Folk Theories of Social Feeds},
year = {2016},
isbn = {9781450333627},
publisher = {Association for Computing Machinery},
address = {New York, NY, USA},
url = {https://doi.org/10.1145/2858036.2858494},
doi = {10.1145/2858036.2858494},
booktitle = {Proceedings of the 2016 CHI Conference on Human Factors in Computing Systems},
pages = {2371–2382},
numpages = {12},
location = {San Jose, California, USA},
series = {CHI '16}
}

@inproceedings{xiao_influence_2025,
author = {Xiao, Qing and Zheng, Yuhang and Fan, Xianzhe and Zhang, Bingbing and Lu, Zhicong},
title = {Let's Influence Algorithms Together: How Millions of Fans Build Collective Understanding of Algorithms and Organize Coordinated Algorithmic Actions},
year = {2025},
isbn = {9798400713941},
publisher = {Association for Computing Machinery},
address = {New York, NY, USA},
url = {https://doi.org/10.1145/3706598.3713279},
doi = {10.1145/3706598.3713279},
booktitle = {Proceedings of the 2025 CHI Conference on Human Factors in Computing Systems},
articleno = {1115},
numpages = {19},
location = {
},
series = {CHI '25}
}

@inproceedings{devito_algorithms_2017,
author = {DeVito, Michael Ann and Gergle, Darren and Birnholtz, Jeremy},
title = {"Algorithms ruin everything": \#RIPTwitter, Folk Theories, and Resistance to Algorithmic Change in Social Media},
year = {2017},
isbn = {9781450346559},
publisher = {Association for Computing Machinery},
address = {New York, NY, USA},
url = {https://doi.org/10.1145/3025453.3025659},
doi = {10.1145/3025453.3025659},
booktitle = {Proceedings of the 2017 CHI Conference on Human Factors in Computing Systems},
pages = {3163–3174},
numpages = {12},
location = {Denver, Colorado, USA},
series = {CHI '17}
}

@inproceedings{li_beyond_2025,
author = {Li, Wenqi and Kuo, Jui-Ching and Sheng, Manyu and Zhang, Pengyi and Wu, Qunfang},
title = {Beyond Explicit and Implicit: How Users Provide Feedback to Shape Personalized Recommendation Content},
year = {2025},
isbn = {9798400713941},
publisher = {Association for Computing Machinery},
address = {New York, NY, USA},
url = {https://doi.org/10.1145/3706598.3713241},
doi = {10.1145/3706598.3713241},
booktitle = {Proceedings of the 2025 CHI Conference on Human Factors in Computing Systems},
articleno = {893},
numpages = {17},
location = {
},
series = {CHI '25}
}

@inproceedings{oard1998implicit,
  title={Implicit feedback for recommender systems},
  author={Oard, Douglas W and Kim, Jinmook and others},
  booktitle={Proceedings of the AAAI workshop on recommender systems},
  volume={83},
  pages={81--83},
  year={1998},
  organization={Madison, WI}
}

@Article{Widder2024,
author={Widder, David Gray
and Whittaker, Meredith
and West, Sarah Myers},
title={Why `open' AI systems are actually closed, and why this matters},
journal={Nature},
year={2024},
month={Nov},
day={01},
volume={635},
number={8040},
pages={827-833},
issn={1476-4687},
doi={10.1038/s41586-024-08141-1},
url={https://doi.org/10.1038/s41586-024-08141-1}
}

@article{Boenig-Liptsin02092022,
author = {Margarita Boenig-Liptsin and Anissa Tanweer and Ari Edmundson},
title = {Data Science Ethos Lifecycle: Interplay of Ethical Thinking and Data Science Practice},
journal = {Journal of Statistics and Data Science Education},
volume = {30},
number = {3},
pages = {228--240},
year = {2022},
publisher = {Taylor \& Francis},
doi = {10.1080/26939169.2022.2089411},


URL = { 
    
        https://doi.org/10.1080/26939169.2022.2089411
    
    

},
eprint = { 
    
        https://doi.org/10.1080/26939169.2022.2089411
    
    

}

}

@book{friedman2019value,
  title={Value sensitive design: Shaping technology with moral imagination},
  author={Friedman, Batya and Hendry, David G},
  year={2019},
  publisher={Mit Press}
}

@inproceedings{wang_identities_2025,
author = {Wang, Angelina},
title = {Identities are not Interchangeable: The Problem of Overgeneralization in Fair Machine Learning},
year = {2025},
isbn = {9798400714825},
publisher = {Association for Computing Machinery},
address = {New York, NY, USA},
url = {https://doi.org/10.1145/3715275.3732033},
doi = {10.1145/3715275.3732033},
booktitle = {Proceedings of the 2025 ACM Conference on Fairness, Accountability, and Transparency},
pages = {485–497},
numpages = {13},
location = {
},
series = {FAccT '25}
}

@inproceedings{wang_towards_2022,
author = {Wang, Angelina and Ramaswamy, Vikram V and Russakovsky, Olga},
title = {Towards Intersectionality in Machine Learning: Including More Identities, Handling Underrepresentation, and Performing Evaluation},
year = {2022},
isbn = {9781450393522},
publisher = {Association for Computing Machinery},
address = {New York, NY, USA},
url = {https://doi.org/10.1145/3531146.3533101},
doi = {10.1145/3531146.3533101},
booktitle = {Proceedings of the 2022 ACM Conference on Fairness, Accountability, and Transparency},
pages = {336–349},
numpages = {14},
location = {Seoul, Republic of Korea},
series = {FAccT '22}
}

@inproceedings{bender_dangers_2021,
author = {Bender, Emily M. and Gebru, Timnit and McMillan-Major, Angelina and Shmitchell, Shmargaret},
title = {On the Dangers of Stochastic Parrots: Can Language Models Be Too Big?},
year = {2021},
isbn = {9781450383097},
publisher = {Association for Computing Machinery},
address = {New York, NY, USA},
url = {https://doi.org/10.1145/3442188.3445922},
doi = {10.1145/3442188.3445922},
booktitle = {Proceedings of the 2021 ACM Conference on Fairness, Accountability, and Transparency},
pages = {610–623},
numpages = {14},
location = {Virtual Event, Canada},
series = {FAccT '21}
}

@inproceedings{caine_local_2016,
	address = {San Jose California USA},
	title = {Local {Standards} for {Sample} {Size} at {CHI}},
	isbn = {978-1-4503-3362-7},
	url = {https://dl.acm.org/doi/10.1145/2858036.2858498},
	doi = {10.1145/2858036.2858498},
	language = {en},
	urldate = {2024-12-01},
	booktitle = {Proceedings of the 2016 {CHI} {Conference} on {Human} {Factors} in {Computing} {Systems}},
	publisher = {ACM},
	author = {Caine, Kelly},
	month = may,
	year = {2016},
	pages = {981--992},
}

@inproceedings{black_toward_2023,
	address = {Boston MA USA},
	series = {{EAAMO}},
	title = {Toward {Operationalizing} {Pipeline}-aware {ML} {Fairness}: {A} {Research} {Agenda} for {Developing} {Practical} {Guidelines} and {Tools}},
	isbn = {979-8-4007-0381-2},
	shorttitle = {Toward {Operationalizing} {Pipeline}-aware {ML} {Fairness}},
	url = {https://dl.acm.org/doi/10.1145/3617694.3623259},
	doi = {10.1145/3617694.3623259},
	language = {en},
	urldate = {2025-03-26},
	booktitle = {Equity and {Access} in {Algorithms}, {Mechanisms}, and {Optimization}},
	publisher = {ACM},
	author = {Black, Emily and Naidu, Rakshit and Ghani, Rayid and Rodolfa, Kit and Ho, Daniel and Heidari, Hoda},
	month = oct,
	year = {2023},
	pages = {1--11},
}

@inproceedings{suresh_framework_2021,
	address = {-- NY USA},
	series = {{EAAMO}},
	title = {A {Framework} for {Understanding} {Sources} of {Harm} throughout the {Machine} {Learning} {Life} {Cycle}},
	isbn = {978-1-4503-8553-4},
	url = {https://dl.acm.org/doi/10.1145/3465416.3483305},
	doi = {10.1145/3465416.3483305},
	language = {en},
	urldate = {2024-01-26},
	booktitle = {Equity and {Access} in {Algorithms}, {Mechanisms}, and {Optimization}},
	publisher = {ACM},
	author = {Suresh, Harini and Guttag, John},
	month = oct,
	year = {2021},
	pages = {1--9},
}

@article{braun_using_2006,
	title = {Using thematic analysis in psychology},
	volume = {3},
	issn = {1478-0887, 1478-0895},
	url = {http://www.tandfonline.com/doi/abs/10.1191/1478088706qp063oa},
	doi = {10.1191/1478088706qp063oa},
	language = {en},
	number = {2},
	urldate = {2024-07-22},
	journal = {Qualitative Research in Psychology},
	author = {Braun, Virginia and Clarke, Victoria},
	month = jan,
	year = {2006},
	pages = {77--101},
}

@misc{wired_dsa,
url={https://www.wired.com/story/big-tech-companies-in-the-us-have-been-told-not-to-apply-the-digital-services-act/}, 
title={The FTC Warns Big Tech Companies Not to Apply the Digital Services Act},
author={Mila Fiordalisi},
howpublished={Wired},
year={2025},
month={08}
}

@inproceedings{ojewale_2025_towards,
author = {Ojewale, Victor and Steed, Ryan and Vecchione, Briana and Birhane, Abeba and Raji, Inioluwa Deborah},
title = {Towards AI Accountability Infrastructure: Gaps and Opportunities in AI Audit Tooling},
year = {2025},
isbn = {9798400713941},
publisher = {Association for Computing Machinery},
address = {New York, NY, USA},
url = {https://doi.org/10.1145/3706598.3713301},
doi = {10.1145/3706598.3713301},
booktitle = {Proceedings of the 2025 CHI Conference on Human Factors in Computing Systems},
articleno = {815},
numpages = {29},
location = {
},
series = {CHI '25}
}

@misc{moharana2025accessibilitypeopleworkthing,
      title={"Accessibility people, you go work on that thing of yours over there": Addressing Disability Inclusion in AI Product Organizations}, 
      author={Sanika Moharana and Cynthia L. Bennett and Erin Buehler and Michael Madaio and Vinita Tibdewal and Shaun K. Kane},
      year={2025},
      eprint={2508.16607},
      archivePrefix={arXiv},
      primaryClass={cs.HC},
      url={https://arxiv.org/abs/2508.16607}, 
}

@misc{guerdan2025measurementbricolageexaminingdata,
      title={Measurement as Bricolage: Examining How Data Scientists Construct Target Variables for Predictive Modeling Tasks}, 
      author={Luke Guerdan and Devansh Saxena and Stevie Chancellor and Zhiwei Steven Wu and Kenneth Holstein},
      year={2025},
      eprint={2507.02819},
      archivePrefix={arXiv},
      primaryClass={cs.HC},
      url={https://arxiv.org/abs/2507.02819}, 
}

@misc{harvey2024gapsresearchpracticemeasuring,
      title={Gaps Between Research and Practice When Measuring Representational Harms Caused by LLM-Based Systems}, 
      author={Emma Harvey and Emily Sheng and Su Lin Blodgett and Alexandra Chouldechova and Jean Garcia-Gathright and Alexandra Olteanu and Hanna Wallach},
      year={2024},
      eprint={2411.15662},
      archivePrefix={arXiv},
      primaryClass={cs.CY},
      url={https://arxiv.org/abs/2411.15662}, 
}

@inproceedings{harvey-etal-2025-understanding,
    title = "Understanding and Meeting Practitioner Needs When Measuring Representational Harms Caused by {LLM}-Based Systems",
    author = "Harvey, Emma  and
      Sheng, Emily  and
      Blodgett, Su Lin  and
      Chouldechova, Alexandra  and
      Garcia-Gathright, Jean  and
      Olteanu, Alexandra  and
      Wallach, Hanna",
    editor = "Che, Wanxiang  and
      Nabende, Joyce  and
      Shutova, Ekaterina  and
      Pilehvar, Mohammad Taher",
    booktitle = "Findings of the Association for Computational Linguistics: ACL 2025",
    month = jul,
    year = "2025",
    address = "Vienna, Austria",
    publisher = "Association for Computational Linguistics",
    url = "https://aclanthology.org/2025.findings-acl.947/",
    doi = "10.18653/v1/2025.findings-acl.947",
    pages = "18423--18440",
    ISBN = "979-8-89176-256-5",
}

@inproceedings{wright_null_2024,
	address = {Rio de Janeiro Brazil},
	series = {{FAccT}},
	title = {Null {Compliance}: {NYC} {Local} {Law} 144 and the challenges of algorithm accountability},
	isbn = {9798400704505},
	shorttitle = {Null {Compliance}},
	url = {https://dl.acm.org/doi/10.1145/3630106.3658998},
	doi = {10.1145/3630106.3658998},
	language = {en},
	urldate = {2024-06-10},
	booktitle = {The 2024 {ACM} {Conference} on {Fairness}, {Accountability}, and {Transparency}},
	publisher = {ACM},
	author = {Wright, Lucas and Muenster, Roxana Mika and Vecchione, Briana and Qu, Tianyao and Cai, Pika (Senhuang) and Smith, Alan and {Comm 2450 Student Investigators} and Metcalf, Jacob and Matias, J. Nathan},
	month = jun,
	year = {2024},
	pages = {1701--1713},
}

@inproceedings{balayn__2023,
	address = {Montreal QC Canada},
	series = {{AIES}},
	title = {“{Fairness} {Toolkits}, {A} {Checkbox} {Culture}?” {On} the {Factors} that {Fragment} {Developer} {Practices} in {Handling} {Algorithmic} {Harms}},
	isbn = {9798400702310},
	shorttitle = {“{Fairness} {Toolkits}, {A} {Checkbox} {Culture}?},
	url = {https://dl.acm.org/doi/10.1145/3600211.3604674},
	doi = {10.1145/3600211.3604674},
	language = {en},
	urldate = {2023-08-30},
	booktitle = {Proceedings of the 2023 {AAAI}/{ACM} {Conference} on {AI}, {Ethics}, and {Society}},
	publisher = {ACM},
	author = {Balayn, Agathe and Yurrita, Mireia and Yang, Jie and Gadiraju, Ujwal},
	month = aug,
	year = {2023},
	pages = {482--495},
}

@inproceedings{lee_landscape_2021,
	address = {Yokohama Japan},
	series = {{CHI}},
	title = {The {Landscape} and {Gaps} in {Open} {Source} {Fairness} {Toolkits}},
	isbn = {978-1-4503-8096-6},
	url = {https://dl.acm.org/doi/10.1145/3411764.3445261},
	doi = {10.1145/3411764.3445261},
	language = {en},
	urldate = {2023-07-10},
	booktitle = {Proceedings of the 2021 {CHI} {Conference} on {Human} {Factors} in {Computing} {Systems}},
	publisher = {ACM},
	author = {Lee, Michelle Seng Ah and Singh, Jat},
	month = may,
	year = {2021},
	pages = {1--13},
}

@inproceedings{richardson_towards_2021,
	address = {Yokohama Japan},
	title = {Towards {Fairness} in {Practice}: {A} {Practitioner}-{Oriented} {Rubric} for {Evaluating} {Fair} {ML} {Toolkits}},
	isbn = {978-1-4503-8096-6},
	shorttitle = {Towards {Fairness} in {Practice}},
	url = {https://dl.acm.org/doi/10.1145/3411764.3445604},
	doi = {10.1145/3411764.3445604},
	language = {en},
	urldate = {2024-05-28},
	booktitle = {Proceedings of the 2021 {CHI} {Conference} on {Human} {Factors} in {Computing} {Systems}},
	publisher = {ACM},
	author = {Richardson, Brianna and Garcia-Gathright, Jean and Way, Samuel F. and Thom, Jennifer and Cramer, Henriette},
	month = may,
	year = {2021},
	pages = {1--13},
}

@inproceedings{deng_understanding_2023,
	address = {Hamburg Germany},
	title = {Understanding {Practices}, {Challenges}, and {Opportunities} for {User}-{Engaged} {Algorithm} {Auditing} in {Industry} {Practice}},
	isbn = {978-1-4503-9421-5},
	url = {https://dl.acm.org/doi/10.1145/3544548.3581026},
	doi = {10.1145/3544548.3581026},
	language = {en},
	urldate = {2024-06-13},
	booktitle = {Proceedings of the 2023 {CHI} {Conference} on {Human} {Factors} in {Computing} {Systems}},
	publisher = {ACM},
	author = {Deng, Wesley Hanwen and Guo, Boyuan and Devrio, Alicia and Shen, Hong and Eslami, Motahhare and Holstein, Kenneth},
	month = apr,
	year = {2023},
	pages = {1--18},
}

@inproceedings{costanza-chock_who_2022,
	address = {Seoul Republic of Korea},
	series = {{FAccT}},
	title = {Who {Audits} the {Auditors}? {Recommendations} from a field scan of the algorithmic auditing ecosystem},
	isbn = {978-1-4503-9352-2},
	shorttitle = {Who {Audits} the {Auditors}?},
	url = {https://dl.acm.org/doi/10.1145/3531146.3533213},
	doi = {10.1145/3531146.3533213},
	language = {en},
	urldate = {2023-04-07},
	booktitle = {2022 {ACM} {Conference} on {Fairness}, {Accountability}, and {Transparency}},
	publisher = {ACM},
	author = {Costanza-Chock, Sasha and Harvey, Emma and Raji, Inioluwa Deborah and Czernuszenko, Martha and Buolamwini, Joy},
	month = jun,
	year = {2022},
	pages = {1571--1583},
}

@inproceedings{kaur_interpreting_2020,
	address = {Honolulu HI USA},
	series = {{CHI}},
	title = {Interpreting {Interpretability}: {Understanding} {Data} {Scientists}' {Use} of {Interpretability} {Tools} for {Machine} {Learning}},
	isbn = {978-1-4503-6708-0},
	shorttitle = {Interpreting {Interpretability}},
	url = {https://dl.acm.org/doi/10.1145/3313831.3376219},
	doi = {10.1145/3313831.3376219},
	language = {en},
	urldate = {2024-10-14},
	booktitle = {Proceedings of the 2020 {CHI} {Conference} on {Human} {Factors} in {Computing} {Systems}},
	publisher = {ACM},
	author = {Kaur, Harmanpreet and Nori, Harsha and Jenkins, Samuel and Caruana, Rich and Wallach, Hanna and Wortman Vaughan, Jennifer},
	month = apr,
	year = {2020},
	pages = {1--14},
}

@inproceedings{cramer_translation_2019,
	address = {Glasgow Scotland Uk},
	series = {{CHI} {EA}},
	title = {Translation, {Tracks} \& {Data}: an {Algorithmic} {Bias} {Effort} in {Practice}},
	isbn = {978-1-4503-5971-9},
	shorttitle = {Translation, {Tracks} \& {Data}},
	url = {https://dl.acm.org/doi/10.1145/3290607.3299057},
	doi = {10.1145/3290607.3299057},
	language = {en},
	urldate = {2024-10-14},
	booktitle = {Extended {Abstracts} of the 2019 {CHI} {Conference} on {Human} {Factors} in {Computing} {Systems}},
	publisher = {ACM},
	author = {Cramer, Henriette and Garcia-Gathright, Jean and Reddy, Sravana and Springer, Aaron and Takeo Bouyer, Romain},
	month = may,
	year = {2019},
	pages = {1--8},
}

@misc{garcia-gathright_assessing_2018,
	title = {Assessing and {Addressing} {Algorithmic} {Bias} - {But} {Before} {We} {Get} {There}},
	url = {http://arxiv.org/abs/1809.03332},
	language = {en},
	urldate = {2024-10-14},
	publisher = {arXiv},
	author = {Garcia-Gathright, Jean and Springer, Aaron and Cramer, Henriette},
	month = sep,
	year = {2018},
	note = {arXiv:1809.03332 [cs]},
}

@article{morley_operationalising_2023,
	title = {Operationalising {AI} ethics: barriers, enablers and next steps},
	volume = {38},
	issn = {0951-5666, 1435-5655},
	shorttitle = {Operationalising {AI} ethics},
	url = {https://link.springer.com/10.1007/s00146-021-01308-8},
	doi = {10.1007/s00146-021-01308-8},
	language = {en},
	number = {1},
	urldate = {2024-10-16},
	journal = {AI \& SOCIETY},
	author = {Morley, Jessica and Kinsey, Libby and Elhalal, Anat and Garcia, Francesca and Ziosi, Marta and Floridi, Luciano},
	month = feb,
	year = {2023},
	pages = {411--423},
}

@inproceedings{wang_strategies_2024,
	series = {{AIES}},
	title = {Strategies for {Increasing} {Corporate} {Responsible} {AI} {Prioritization}},
	url = {https://ojs.aaai.org/index.php/AIES/article/view/31743},
	language = {en},
	booktitle = {Proceedings of the {Seventh} {AAAI}/{ACM} {Conference} on {AI}, {Ethics}, and {Society} ({AIES2024})},
	author = {Wang, Angelina and Datta, Teresa and Dickerson, John P},
	month = oct,
	year = {2024},
}

@inproceedings{sambasivan_everyone_2021,
	address = {Yokohama Japan},
	title = {“{Everyone} wants to do the model work, not the data work”: {Data} {Cascades} in {High}-{Stakes} {AI}},
	isbn = {978-1-4503-8096-6},
	shorttitle = {“{Everyone} wants to do the model work, not the data work”},
	url = {https://dl.acm.org/doi/10.1145/3411764.3445518},
	doi = {10.1145/3411764.3445518},
	language = {en},
	urldate = {2024-11-04},
	booktitle = {Proceedings of the 2021 {CHI} {Conference} on {Human} {Factors} in {Computing} {Systems}},
	publisher = {ACM},
	author = {Sambasivan, Nithya and Kapania, Shivani and Highfill, Hannah and Akrong, Diana and Paritosh, Praveen and Aroyo, Lora M},
	month = may,
	year = {2021},
	pages = {1--15}}

@inproceedings{grgic-hlaca_case_2016,
	address = {Barcelona, Spain},
	series = {{NIPS}},
	title = {The {Case} for {Process} {Fairness} in {Learning}: {Feature} {Selection} for {Fair} {Decision} {Making}},
	shorttitle = {The {Case} for {Process} {Fairness} in {Learning}},
	url = {https://www.semanticscholar.org/paper/The-Case-for-Process-Fairness-in-Learning%3A-Feature-Grgic-Hlaca-Zafar/fdb6a159cb65f4d1147224998d56e67f0398948b},
	urldate = {2023-04-06},
	booktitle = {Symposium on {Machine} {Learning} and the {Law} at the 29th {Conference} on {Neural} {Information} {Processing} {Systems}},
	author = {Grgic-Hlaca, Nina and Zafar, M. B. and Gummadi, K. and Weller, Adrian},
	month = dec,
	year = {2016},
}

@inproceedings{rieke_imperfect_2022,
	address = {Seoul Republic of Korea},
	series = {{FAccT}},
	title = {Imperfect {Inferences}: {A} {Practical} {Assessment}},
	isbn = {978-1-4503-9352-2},
	shorttitle = {Imperfect {Inferences}},
	url = {https://dl.acm.org/doi/10.1145/3531146.3533140},
	doi = {10.1145/3531146.3533140},
	language = {en},
	urldate = {2023-04-11},
	booktitle = {2022 {ACM} {Conference} on {Fairness}, {Accountability}, and {Transparency}},
	publisher = {ACM},
	author = {Rieke, Aaron and Southerland, Vincent and Svirsky, Dan and Hsu, Mingwei},
	month = jun,
	year = {2022},
	pages = {767--777},
}

@inproceedings{quinonero_candela_disentangling_2023,
	address = {New York, NY, USA},
	series = {{FAccT}},
	title = {Disentangling and {Operationalizing} {AI} {Fairness} at {LinkedIn}},
	isbn = {9798400701924},
	url = {https://dl.acm.org/doi/10.1145/3593013.3594075},
	doi = {10.1145/3593013.3594075},
	urldate = {2023-06-13},
	booktitle = {Proceedings of the 2023 {ACM} {Conference} on {Fairness}, {Accountability}, and {Transparency}},
	publisher = {Association for Computing Machinery},
	author = {Quiñonero Candela, Joaquin and Wu, Yuwen and Hsu, Brian and Jain, Sakshi and Ramos, Jennifer and Adams, Jon and Hallman, Robert and Basu, Kinjal},
	month = jun,
	year = {2023},
	pages = {1213--1228},
}

@article{robertson_not_2022,
	series = {{CSCW}},
	title = {Not {Another} {School} {Resource} {Map}: {Meeting} {Underserved} {Families}' {Information} {Needs} {Requires} {Trusting} {Relationships} and {Personalized} {Care}},
	volume = {6},
	issn = {2573-0142},
	shorttitle = {Not {Another} {School} {Resource} {Map}},
	url = {https://dl.acm.org/doi/10.1145/3555207},
	doi = {10.1145/3555207},
	language = {en},
	number = {CSCW2},
	urldate = {2023-09-26},
	journal = {Proceedings of the ACM on Human-Computer Interaction},
	author = {Robertson, Samantha and Nguyen, Tonya and Salehi, Niloufar},
	month = nov,
	year = {2022},
	pages = {1--23},
}

@article{metcalf_owning_2019,
	title = {Owning {Ethics}: {Corporate} {Logics}, {Silicon} {Valley}, and the {Institutionalization} of {Ethics}},
	volume = {86},
	issn = {1944-768X},
	shorttitle = {Owning {Ethics}},
	url = {https://muse.jhu.edu/article/732185},
	doi = {10.1353/sor.2019.0022},
	language = {en},
	number = {2},
	urldate = {2023-10-25},
	journal = {Social Research: An International Quarterly},
	author = {Metcalf, Jacob and Moss, Emanuel and Boyd, Danah},
	month = jun,
	year = {2019},
	pages = {449--476},
}

@inproceedings{smith_pragmatic_2025,
	address = {Athens Greece},
	series = {{FAccT}},
	title = {Pragmatic {Fairness}: {Evaluating} {ML} {Fairness} {Within} the {Constraints} of {Industry}},
	isbn = {9798400714825},
	shorttitle = {Pragmatic {Fairness}},
	url = {https://dl.acm.org/doi/10.1145/3715275.3732040},
	doi = {10.1145/3715275.3732040},
	language = {en},
	urldate = {2025-07-01},
	booktitle = {Proceedings of the 2025 {ACM} {Conference} on {Fairness}, {Accountability}, and {Transparency}},
	publisher = {ACM},
	author = {Smith, Jessie J. and Madaio, Michael and Burke, Robin and Fiesler, Casey},
	month = jun,
	year = {2025},
	pages = {628--638},
}

@article{feng_has_2022,
	series = {{AAAI}},
	title = {Has {CEO} {Gender} {Bias} {Really} {Been} {Fixed}? {Adversarial} {Attacking} and {Improving} {Gender} {Fairness} in {Image} {Search}},
	volume = {36},
	copyright = {Copyright (c) 2022 Association for the Advancement of Artificial Intelligence},
	issn = {2374-3468},
	shorttitle = {Has {CEO} {Gender} {Bias} {Really} {Been} {Fixed}?},
	url = {https://ojs.aaai.org/index.php/AAAI/article/view/21445},
	doi = {10.1609/aaai.v36i11.21445},
	language = {en},
	number = {11},
	urldate = {2023-04-07},
	journal = {Proceedings of the AAAI Conference on Artificial Intelligence},
	author = {Feng, Yunhe and Shah, Chirag},
	month = jun,
	year = {2022},
	pages = {11882--11890},
}

@article{sweeney_discrimination_2013,
	title = {Discrimination in online ad delivery},
	volume = {56},
	issn = {0001-0782, 1557-7317},
	url = {https://dl.acm.org/doi/10.1145/2447976.2447990},
	doi = {10.1145/2447976.2447990},
	language = {en},
	number = {5},
	urldate = {2023-10-25},
	journal = {Communications of the ACM},
	author = {Sweeney, Latanya},
	month = may,
	year = {2013},
	pages = {44--54},
}

@misc{markupAmazon, 
url={https://themarkup.org/amazons-advantage/2023/09/28/amazon-ranks-its-own-products-first-ftc-lawsuit-says}, 
title={Amazon Ranks Its Own Products First, FTC Lawsuit Says},
author={Ryan Tate},
year={2023},
month={09},
howpublished={The Markup}}

@inproceedings{ye2025auditing,
author = {Ye, Jinyi and Luceri, Luca and Ferrara, Emilio},
title = {Auditing Political Exposure Bias: Algorithmic Amplification on Twitter/X During the 2024 U.S. Presidential Election},
year = {2025},
isbn = {9798400714825},
publisher = {Association for Computing Machinery},
address = {New York, NY, USA},
url = {https://doi.org/10.1145/3715275.3732159},
doi = {10.1145/3715275.3732159},
booktitle = {Proceedings of the 2025 ACM Conference on Fairness, Accountability, and Transparency},
pages = {2349–2362},
numpages = {14},
location = {
},
series = {FAccT '25}
}

@ARTICLE{anandhan2018social,
  author={Anandhan, Anitha and Shuib, Liyana and Ismail, Maizatul Akmar and Mujtaba, Ghulam},
  journal={IEEE Access}, 
  title={Social Media Recommender Systems: Review and Open Research Issues}, 
  year={2018},
  volume={6},
  number={},
  pages={15608-15628},
  doi={10.1109/ACCESS.2018.2810062}}

@article{huszar2022algorithmic,
author = {Ferenc Huszár  and Sofia Ira Ktena  and Conor O’Brien  and Luca Belli  and Andrew Schlaikjer  and Moritz Hardt },
title = {Algorithmic amplification of politics on Twitter},
journal = {Proceedings of the National Academy of Sciences},
volume = {119},
number = {1},
pages = {e2025334119},
year = {2022},
URL = {https://www.pnas.org/doi/abs/10.1073/pnas.2025334119}}

@article{stray2022building,
   title={Building human values into recommender systems: An interdisciplinary synthesis},
  author={Stray, Jonathan and Halevy, Alon and Assar, Parisa and Hadfield-Menell, Dylan and Boutilier, Craig and Ashar, Amar and Bakalar, Chloe and Beattie, Lex and Ekstrand, Michael and Leibowicz, Claire and others},
  journal={ACM Transactions on Recommender Systems},
  volume={2},
  number={3},
  pages={1--57},
  year={2024},
  publisher={ACM New York, NY}
}

@article{lada2021machine,
  title={How machine learning powers Facebook’s News Feed ranking algorithm},
  author={Lada, Akos and Wang, Meihong and Yan, Tak},
  journal={Facebook Engineering},
  year={2021}
}

@article{belli2020privacy,
  title={Privacy-Aware Recommender Systems Challenge on Twitter's Home Timeline},
  author={Belli, Luca and Ktena, Sofia Ira and Tejani, Alykhan and Lung-Yut-Fon, Alexandre and Portman, Frank and Zhu, Xiao and Xie, Yuanpu and Gupta, Akshay and Bronstein, Michael and Deli{\'c}, Amra and others},
  journal={arXiv preprint arXiv:2004.13715},
  year={2020}
}

@article{gretzel2006persuasion,
  title={Persuasion in recommender systems},
  author={Gretzel, Ulrike and Fesenmaier, Daniel R},
  journal={International Journal of Electronic Commerce},
  volume={11},
  number={2},
  pages={81--100},
  year={2006},
  publisher={Taylor \& Francis}
}

@inproceedings{zalmout2021all,
  title={All You Need to Know to Build a Product Knowledge Graph},
  author={Zalmout, Nasser and Zhang, Chenwei and Li, Xian and Liang, Yan and Dong, Xin Luna},
  booktitle={Proceedings of the 27th ACM SIGKDD Conference on Knowledge Discovery \& Data Mining},
  pages={4090--4091},
  year={2021}
}

@article{chen2020bias,
  title={Bias and debias in recommender system: A survey and future directions},
  author={Chen, Jiawei and Dong, Hande and Wang, Xiang and Feng, Fuli and Wang, Meng and He, Xiangnan},
  journal={ACM Transactions on Information Systems},
  volume={41},
  number={3},
  pages={1--39},
  year={2023},
  publisher={ACM New York, NY}
}

@article{Kanubala_Valera_2025, title={On the Misalignment Between Legal Notions and Statistical Metrics of Intersectional Fairness}, volume={8}, url={https://ojs.aaai.org/index.php/AIES/article/view/36637}, DOI={10.1609/aies.v8i2.36637}, abstractNote={Intersectional (un)fairness, as conceptualized in legal and social theory, emphasizes the non-additive and structurally complex nature of discrimination against individuals at the intersection of multiple sensitive attributes (such as race, gender, etc). Recent works have proposed statistical metrics for intersectional fairness by estimating disparities across groups of individuals sharing two or more sensitive attributes. However, it is unclear if these metrics detect uniquely intersectional discrimination. We therefore pose the following question, Do current statistical intersectional metrics detect the non-additive discrimination highlighted by intersectionality theory? More specifically, to answer this, we run controlled synthetic data experiments that explicitly allow us to control for single, multiple, intersectional, and compounded forms of discrimination. Our analyses show that current statistical metrics for intersectional fairness behave more like multi-attribute disparity measures. Specifically, they respond more strongly to additive or compounded biases than to non-additive interaction effects. While they effectively capture disparities across multiple sensitive attributes, they often fail to detect uniquely intersectional discrimination. These findings reveal a fundamental misalignment between existing intersectional fairness metrics and the legal and theoretical foundations of intersectionality. We argue that if intersectional fairness metrics are to be deemed truly intersectional, they must be explicitly designed to account for the structural, non-additive nature of intersectional discrimination.}, number={2}, journal={Proceedings of the AAAI/ACM Conference on AI, Ethics, and Society}, author={Kanubala, Deborah Dormah and Valera, Isabel}, year={2025}, month={Oct.}, pages={1363-1374} }

@article{ganesh_wild_2025,
	series = {{AIES}},
	title = {The ‘{Wild} {West}’ of {Medicine}: {Exploring} the {Emergence} of ‘{Grassroots}’ {AI} {Governance} in {Radiology}},
	volume = {8},
	issn = {3065-8365},
	shorttitle = {The ‘{Wild} {West}’ of {Medicine}},
	url = {https://ojs.aaai.org/index.php/AIES/article/view/36608},
	doi = {10.1609/aies.v8i2.36608},
	number = {2},
	urldate = {2025-10-24},
	journal = {Proceedings of the AAAI/ACM Conference on AI, Ethics, and Society},
	author = {Ganesh, Bhargavi and Schiff, Daniel S. and Anderson, Stuart},
	month = oct,
	year = {2025},
	pages = {1018--1031},
}

@inproceedings{zemel2013learning,
  title={Learning fair representations},
  author={Zemel, Rich and Wu, Yu and Swersky, Kevin and Pitassi, Toni and Dwork, Cynthia},
  booktitle={International Conference on Machine Learning},
  pages={325--333},
  year={2013}
}

@article{kleinberg2016inherent,
  doi = {10.4230/LIPICS.ITCS.2017.43},
  url = {https://drops.dagstuhl.de/entities/document/10.4230/LIPIcs.ITCS.2017.43},
  author = {Kleinberg,  Jon and Mullainathan,  Sendhil and Raghavan,  Manish},
  language = {en},
  title = {Inherent Trade-Offs in the Fair Determination of Risk Scores},
  publisher = {Schloss Dagstuhl – Leibniz-Zentrum f\"{u}r Informatik},
  year = {2017},
  journal={Leibniz International Proceedings in Informatics},
  copyright = {Creative Commons Attribution 3.0 Unported license}
}

@article{chouldechova2017fair,
  title={Fair prediction with disparate impact: A study of bias in recidivism prediction instruments},
  author={Chouldechova, Alexandra},
  journal={Big data},
  volume={5},
  number={2},
  pages={153--163},
  year={2017},
  publisher={Mary Ann Liebert, Inc. 140 Huguenot Street, 3rd Floor New Rochelle, NY 10801 USA}
}

@inproceedings{dwork2012fairness,
  title={Fairness through awareness},
  author={Dwork, Cynthia and Hardt, Moritz and Pitassi, Toniann and Reingold, Omer and Zemel, Richard},
  booktitle={Proceedings of the 3rd innovations in theoretical computer science conference},
  pages={214--226},
  year={2012},
  organization={ACM}
}

@article{barocas2016big,
  title={Big data's disparate impact},
  author={Barocas, Solon and Selbst, Andrew D},
  journal={Calif. L. Rev.},
  volume={104},
  pages={671},
  year={2016},
  publisher={HeinOnline}
}

@inproceedings{kay2015unequal,
  title={Unequal representation and gender stereotypes in image search results for occupations},
  author={Kay, Matthew and Matuszek, Cynthia and Munson, Sean A},
  booktitle={Proceedings of the 33rd Annual ACM Conference on Human Factors in Computing Systems},
  pages={3819--3828},
  year={2015},
  organization={ACM}
}

@inproceedings{holstein2019improving,
  title={Improving fairness in machine learning systems: What do industry practitioners need?},
  author={Holstein, Kenneth and Wortman Vaughan, Jennifer and Daum{\'e} III, Hal and Dudik, Miro and Wallach, Hanna},
  booktitle={Proceedings of the 2019 CHI Conference on Human Factors in Computing Systems},
  pages={600},
  year={2019},
  organization={ACM}
}

@inproceedings{veale2018fairness,
  title={Fairness and accountability design needs for algorithmic support in high-stakes public sector decision-making},
  author={Veale, Michael and Van Kleek, Max and Binns, Reuben},
  booktitle={Proceedings of the 2018 chi conference on human factors in computing systems},
  pages={440},
  year={2018},
  organization={ACM}
}

@inproceedings{beutel2019putting,
  title={Putting fairness principles into practice: Challenges, metrics, and improvements},
  author={Beutel, Alex and Chen, Jilin and Doshi, Tulsee and Qian, Hai and Woodruff, Allison and Luu, Christine and Kreitmann, Pierre and Bischof, Jonathan and Chi, Ed H},
  booktitle={Proceedings of the 2019 AAAI/ACM Conference on AI, Ethics, and Society},
  pages={453--459},
  year={2019}
}

@inproceedings{madaio2020co,
  title={Co-Designing Checklists to Understand Organizational Challenges and Opportunities around Fairness in AI},
  author={Madaio, Michael A and Stark, Luke and Wortman Vaughan, Jennifer and Wallach, Hanna},
  booktitle={Proceedings of the 2020 CHI Conference on Human Factors in Computing Systems},
  pages={1--14},
  year={2020}
}

@inproceedings{smith2022real,
  title={REAL ML: Recognizing, Exploring, and Articulating Limitations of Machine Learning Research},
  author={Smith, Jessie J and Amershi, Saleema and Barocas, Solon and Wallach, Hanna and Wortman Vaughan, Jennifer},
  booktitle={2022 ACM Conference on Fairness, Accountability, and Transparency},
  pages={587--597},
  year={2022}
}

@article{heger2022understanding,
  title={Understanding machine learning practitioners' data documentation perceptions, needs, challenges, and desiderata},
  author={Heger, Amy K and Marquis, Liz B and Vorvoreanu, Mihaela and Wallach, Hanna and Wortman Vaughan, Jennifer},
  journal={Proceedings of the ACM on Human-Computer Interaction},
  volume={6},
  number={CSCW2},
  pages={1--29},
  year={2022},
  publisher={ACM New York, NY, USA}
}

@article{madaio2022assessing,
  title={Assessing the Fairness of AI Systems: AI Practitioners' Processes, Challenges, and Needs for Support},
  author={Madaio, Michael and Egede, Lisa and Subramonyam, Hariharan and Wortman Vaughan, Jennifer and Wallach, Hanna},
  journal={Proceedings of the ACM on Human-Computer Interaction},
  volume={6},
  number={CSCW1},
  pages={1--26},
  year={2022},
  publisher={ACM New York, NY, USA}
}

@article{wang2022trustworthy,
  title={Trustworthy recommender systems},
  author={Wang, Shoujin and Zhang, Xiuzhen and Wang, Yan and Ricci, Francesco},
  journal={ACM Transactions on Intelligent Systems and Technology},
  volume={15},
  number={4},
  pages={1--20},
  year={2024},
  publisher={ACM New York, NY}
}

@inproceedings{leonhardt2018user,
  title={User fairness in recommender systems},
  author={Leonhardt, Jurek and Anand, Avishek and Khosla, Megha},
  booktitle={Companion Proceedings of the The Web Conference 2018},
  pages={101--102},
  year={2018}
}

@article{abdollahpouri2019multi,
  title={Multi-stakeholder recommendation and its connection to multi-sided fairness},
  author={Abdollahpouri, Himan and Burke, Robin},
  journal={arXiv preprint arXiv:1907.13158},
  year={2019}
}

@inproceedings{mehrotra2018towards,
  title={Towards a fair marketplace: Counterfactual evaluation of the trade-off between relevance, fairness \& satisfaction in recommendation systems},
  author={Mehrotra, Rishabh and McInerney, James and Bouchard, Hugues and Lalmas, Mounia and Diaz, Fernando},
  booktitle={Proceedings of the 27th acm international conference on information and knowledge management},
  pages={2243--2251},
  year={2018}
}

@inproceedings{li2021tutorial,
  title={Tutorial on fairness of machine learning in recommender systems},
  author={Li, Yunqi and Ge, Yingqiang and Zhang, Yongfeng},
  booktitle={Proceedings of the 44th International ACM SIGIR Conference on Research and Development in Information Retrieval},
  pages={2654--2657},
  year={2021}
}

@inproceedings{burke2018balanced,
  title={Balanced neighborhoods for multi-sided fairness in recommendation},
  author={Burke, Robin and Sonboli, Nasim and Ordonez-Gauger, Aldo},
  booktitle={Conference on fairness, accountability and transparency},
  pages={202--214},
  year={2018},
  organization={PMLR}
}

@article{keller2021amplification,
  title={Amplification and its discontents: Why regulating the reach of online content is hard},
  author={Keller, Daphne},
  journal={J. FREE SPEECH L.},
  volume={1},
  pages={227--268},
  year={2021}
}

@inproceedings{lum2022biasing,
  title={De-biasing “bias” measurement},
  author={Lum, Kristian and Zhang, Yunfeng and Bower, Amanda},
  booktitle={Proceedings of the 2022 ACM Conference on Fairness, Accountability, and Transparency},
  pages={379--389},
  year={2022}
}

@article{burke2017multisided,
  title={Multisided fairness for recommendation},
  author={Burke, Robin},
  journal={arXiv preprint arXiv:1707.00093},
  year={2017}
}

@inproceedings{beutel2018latent,
  title={Latent cross: Making use of context in recurrent recommender systems},
  author={Beutel, Alex and Covington, Paul and Jain, Sagar and Xu, Can and Li, Jia and Gatto, Vince and Chi, Ed H},
  booktitle={Proceedings of the Eleventh ACM International Conference on Web Search and Data Mining},
  pages={46--54},
  year={2018}
}

@incollection{ekstrand2022fairness,
  title={Fairness in recommender systems},
  author={Ekstrand, Michael D and Das, Anubrata and Burke, Robin and Diaz, Fernando},
  booktitle={Recommender systems handbook},
  pages={679--707},
  year={2022},
  publisher={Springer}
}

@article{madiega2020digital,
  title={Digital services act},
  author={Madiega, Tambiama},
  journal={European Parliamentary Research Service, PE},
  year={2020}
}

@inproceedings{passi2019problem,
  title={Problem formulation and fairness},
  author={Passi, Samir and Barocas, Solon},
  booktitle={Proceedings of the conference on fairness, accountability, and transparency},
  pages={39--48},
  year={2019}
}

@article{boyd2021datasheets,
author = {Boyd, Karen L.},
title = {Datasheets for Datasets help ML Engineers Notice and Understand Ethical Issues in Training Data},
year = {2021},
issue_date = {October 2021},
publisher = {Association for Computing Machinery},
address = {New York, NY, USA},
volume = {5},
number = {CSCW2},
url = {https://doi.org/10.1145/3479582},
doi = {10.1145/3479582},
journal = {Proc. ACM Hum.-Comput. Interact.},
month = oct,
articleno = {438},
numpages = {27}
}

@article{boyd_adapting_2021,
author = {Boyd, Karen L. and Shilton, Katie},
title = {Adapting Ethical Sensitivity as a Construct to Study Technology Design Teams},
year = {2021},
issue_date = {July 2021},
publisher = {Association for Computing Machinery},
address = {New York, NY, USA},
volume = {5},
number = {GROUP},
url = {https://doi.org/10.1145/3463929},
doi = {10.1145/3463929},
journal = {Proc. ACM Hum.-Comput. Interact.},
month = jul,
articleno = {217},
numpages = {29}
}

@inproceedings{papakyriakopoulos2023augmented,
  title={Augmented datasheets for speech datasets and ethical decision-making},
  author={Papakyriakopoulos, Orestis and Choi, Anna Seo Gyeong and Thong, William and Zhao, Dora and Andrews, Jerone and Bourke, Rebecca and Xiang, Alice and Koenecke, Allison},
  booktitle={Proceedings of the 2023 ACM Conference on Fairness, Accountability, and Transparency},
  pages={881--904},
  year={2023}
}

@article{rakova2021responsible,
  title={Where responsible AI meets reality: Practitioner perspectives on enablers for shifting organizational practices},
  author={Rakova, Bogdana and Yang, Jingying and Cramer, Henriette and Chowdhury, Rumman},
  journal={Proceedings of the ACM on Human-Computer Interaction},
  volume={5},
  number={CSCW1},
  pages={1--23},
  year={2021},
  publisher={ACM New York, NY, USA}
}

@inproceedings{kearns2019empirical,
  title={An empirical study of rich subgroup fairness for machine learning},
  author={Kearns, Michael and Neel, Seth and Roth, Aaron and Wu, Zhiwei Steven},
  booktitle={Proceedings of the conference on fairness, accountability, and transparency},
  pages={100--109},
  year={2019}
}

@inproceedings{pang2023auditing,
  doi = {10.1145/3593013.3594098},
  url = {https://doi.org/10.1145/3593013.3594098},
  year = {2023},
  month = jun,
  publisher = {{ACM}},
  author = {Rock Yuren Pang and Jack Cenatempo and Franklyn Graham and Bridgette Kuehn and Maddy Whisenant and Portia Botchway and Katie Stone Perez and Allison Koenecke},
  title = {Auditing Cross-Cultural Consistency of Human-Annotated Labels for Recommendation Systems},
  booktitle = {2023 {ACM} Conference on Fairness,  Accountability,  and Transparency}
}

@article{Corpus2025,
  title = {As Government Outsources More IT,  Highly Skilled In-House Technologists Are More Essential},
  volume = {68},
  ISSN = {1557-7317},
  url = {http://dx.doi.org/10.1145/3727635},
  DOI = {10.1145/3727635},
  number = {7},
  journal = {Communications of the ACM},
  publisher = {Association for Computing Machinery (ACM)},
  author = {Corpus,  Isabel and Giannella,  Eric and Koenecke,  Allison and Moynihan,  Don},
  year = {2025},
  month = jun,
  pages = {37–40}
}

@article{Alkemade2023,
  title = {Datasheets for Digital Cultural Heritage Datasets},
  volume = {9},
  ISSN = {2059-481X},
  url = {http://dx.doi.org/10.5334/johd.124},
  DOI = {10.5334/johd.124},
  journal = {Journal of Open Humanities Data},
  publisher = {Ubiquity Press,  Ltd.},
  author = {Alkemade,  Henk and Claeyssens,  Steven and Colavizza,  Giovanni and Freire,  Nuno and Lehmann,  J\"{o}rg and Neudecker,  Clemens and Osti,  Giulia and van Strien,  Daniel},
  year = {2023}
}

@article{koenecke2023popular,
  doi = {10.1609/icwsm.v17i1.22163},
  url = {https://doi.org/10.1609/icwsm.v17i1.22163},
  year = {2023},
  month = jun,
  publisher = {Association for the Advancement of Artificial Intelligence ({AAAI})},
  volume = {17},
  pages = {494--506},
  author = {Allison Koenecke and Eric Giannella and Robb Willer and Sharad Goel},
  title = {Popular Support for Balancing Equity and Efficiency in Resource Allocation: A Case Study in Online Advertising to Increase Welfare Program Awareness},
  journal = {Proceedings of the International {AAAI} Conference on Web and Social Media}
}

@inproceedings{hutchinson201950,
  title={50 years of test (un) fairness: Lessons for machine learning},
  author={Hutchinson, Ben and Mitchell, Margaret},
  booktitle={Proceedings of the conference on fairness, accountability, and transparency},
  pages={49--58},
  year={2019}
}

@article{corbett2023measure,
  title={The measure and mismeasure of fairness},
  author={Corbett-Davies, Sam and Gaebler, Johann D and Nilforoshan, Hamed and Shroff, Ravi and Goel, Sharad},
  journal={Journal of Machine Learning Research},
  volume={24},
  number={312},
  pages={1--117},
  year={2023}
}

@inproceedings{ali2023walking,
  title={Walking the walk of AI ethics: Organizational challenges and the individualization of risk among ethics entrepreneurs},
  author={Ali, Sanna J and Christin, Ang{\`e}le and Smart, Andrew and Katila, Riitta},
  booktitle={Proceedings of the 2023 ACM Conference on Fairness, Accountability, and Transparency},
  pages={217--226},
  year={2023}
}

@inproceedings{sonboli2021fairness,
  title={Fairness and transparency in recommendation: The users’ perspective},
  author={Sonboli, Nasim and Smith, Jessie J and Cabral Berenfus, Florencia and Burke, Robin and Fiesler, Casey},
  booktitle={Proceedings of the 29th ACM Conference on User Modeling, Adaptation and Personalization},
  pages={274--279},
  year={2021}
}

@article{abdollahpouri2019managing,
  title={Managing popularity bias in recommender systems with personalized re-ranking},
  author={Abdollahpouri, Himan and Burke, Robin and Mobasher, Bamshad},
  journal={arXiv preprint arXiv:1901.07555},
  year={2019}
}

@inproceedings{smith2023scoping,
  title={Scoping fairness objectives and identifying fairness metrics for recommender systems: The practitioners’ perspective},
  author={Smith, Jessie J and Beattie, Lex and Cramer, Henriette},
  booktitle={Proceedings of the ACM Web Conference 2023},
  pages={3648--3659},
  year={2023}
}

@inproceedings{smith2023many,
  title={The many faces of fairness: Exploring the institutional logics of multistakeholder microlending recommendation},
  author={Smith, Jessie J and Buhayh, Anas and Kathait, Anushka and Ragothaman, Pradeep and Mattei, Nicholas and Burke, Robin and Voida, Amy},
  booktitle={Proceedings of the 2023 ACM Conference on Fairness, Accountability, and Transparency},
  pages={1652--1663},
  year={2023}
}

@inproceedings{scheuerman2024walled,
  title={In the walled garden: Challenges and opportunities for research on the practices of the AI tech industry},
  author={Scheuerman, Morgan Klaus},
  booktitle={Proceedings of the 2024 ACM Conference on Fairness, Accountability, and Transparency},
  pages={456--466},
  year={2024}
}

@inproceedings{deng2022exploring,
  title={Exploring how machine learning practitioners (try to) use fairness toolkits},
  author={Deng, Wesley Hanwen and Nagireddy, Manish and Lee, Michelle Seng Ah and Singh, Jatinder and Wu, Zhiwei Steven and Holstein, Kenneth and Zhu, Haiyi},
  booktitle={Proceedings of the 2022 ACM Conference on Fairness, Accountability, and Transparency},
  pages={473--484},
  year={2022}
}

@inproceedings{deng2023investigating,
  title={Investigating practices and opportunities for cross-functional collaboration around AI fairness in industry practice},
  author={Deng, Wesley Hanwen and Yildirim, Nur and Chang, Monica and Eslami, Motahhare and Holstein, Kenneth and Madaio, Michael},
  booktitle={Proceedings of the 2023 ACM Conference on Fairness, Accountability, and Transparency},
  pages={705--716},
  year={2023}
}

@inproceedings{smith2024recommend,
  title={Recommend me? designing fairness metrics with providers},
  author={Smith, Jessie J and Satwani, Aishwarya and Burke, Robin and Fiesler, Casey},
  booktitle={Proceedings of the 2024 ACM Conference on Fairness, Accountability, and Transparency},
  pages={2389--2399},
  year={2024}
}

\appendix
\newpage
\section{Semi-Structured Interview Guide}
\begin{table*}[ht]
\begin{small}
\begin{tabularx}{\textwidth}{llX}
\toprule
\textbf{Interview Phase} & \textbf{Question Type} & \textbf{Questions} \\
\midrule
\multirow{3}{*}{Ice-Breaking} & 
        Participant background & Please tell me a bit more about yourself. \\ \cmidrule{2-3}
        & Participant company & Please tell me something about the recommender system that you are helping to build. \\ \cmidrule{2-3}
        & \multirow{3}{*}{Participant experience} & 
        How long have you been working in this field?\\
        & & How long you have been working at your current company?\\
        & & Are you focusing on research, production, or a mixture of both?
        \\
\midrule
    \multirow{16}{*}{\shortstack[l]{Workflow and \\Approach to Fairness}} & \multirow{4}{*}{\shortstack[l]{Participant Workflow \\ (opening questions)}} & Could you describe a recent project's workflow? \\
     &  & Are there external teams that you interact with regularly? \\
     &  & Could you categorize your basic daily stages? \\
     &  & How do you feel about your workflow? \\ \cmidrule{2-3}
     & \multirow{3}{*}{\shortstack[l]{Participant Workflow \\ (follow-up questions)}} & 
     Which stage do you think you spend most of your time on?\\
     &  & Would you mind sharing a recent interaction with [X] team?\\
     &  & How do you feel while interacting with [X] team? \\ \cmidrule{2-3}
     & \multirow{6}{*}{Participant Approach to Fairness} & How do you define bias in your daily work?\\ 
     & & Where do you find metrics and tools to quantify and detect bias? \\
     &  & Can you share a recent experience detecting bias in the RS that you work with?\\
     &  &  Do you rely on a taxonomy of unfairness or harm in your work? If yes, how do you find it useful? \\
     &  & What amount of time do you spend on fairness in your daily work? \\
     &  & How do you feel about your experience with fairness in RS? \\ \cmidrule{2-3}
     & \multirow{5}{*}{Organizational Approach to Fairness} & Is fairness in your organization a team effort, or is responsibility for fairness distributed just to one person or a small group? What do you think of this assignment?\\
     &  & In terms of mitigation approaches, do you prefer model-centric ones or data-centric ones? Why? \\
     &  & Could you describe a recent time that improving fairness failed or was extremely difficult? \\
     &  & What are the barriers you find that will prevent you from improving fairness? \\
     &  & How do you like the current partition of the time? Would you prefer to spend more or less effort on fairness? Why? \\ 
\midrule
\multirow{2}{*}{Work Experience} & 
        \multirow{2}{*}{Participant Preferences} & Describe fairness-related tools that you are familiar with. Which tools are you more or less likely to use? Why?\\ 
        & & Can you rate your professional experiences in RS in terms of emotional support, work intensity, and stress level?\\
\bottomrule
\end{tabularx}
\end{small}
\caption{\textbf{Semi-Structured Interview Guide.} As is typical of semi-structured interview studies, not all participants were asked the exact same set of questions in the same order. Some participants were asked additional follow-up questions as appropriate.}\label{tab:example:more_question}
\Description{A semi-structured interview guide. It includes: (1) ice-breaking questions: Please tell me a bit more about yourself, Please tell me something about the recommender system that you are helping to build, How long have you been working in this field?, How long you have been working at your current company?, Are you focusing on research, production, or a mixture of both? (2) Questions about participants' workflows and approaches to fairness:  Could you describe a recent project's workflow?, Are there external teams that you interact with regularly?, Could you categorize your basic daily stages?, How do you feel about your workflow?, Which stage do you think you spend most of your time on?, Would you mind sharing a recent interaction with [X] team?, How do you feel while interacting with [X] team?, How do you define bias in your daily work?, Where do you find metrics and tools to quantify and detect bias?, Can you share a recent experience detecting bias in the RS that you work with?, Do you rely on a taxonomy of unfairness or harm in your work? If yes, how do you find it useful?, What amount of time do you spend on fairness in your daily work?, How do you feel about your experience with fairness in RS?, Is fairness in your organization a team effort, or is responsibility for fairness distributed just to one person or a small group? What do you think of this assignment?, In terms of mitigation approaches, do you prefer model-centric ones or data-centric ones? Why?, Could you describe a recent time that improving fairness failed or was extremely difficult?, What are the barriers you find that will prevent you from improving fairness?, How do you like the current partition of the time? Would you prefer to spend more or less effort on fairness? Why? (3) Work experience questions: Describe fairness-related tools that you are familiar with. Which tools are you more or less likely to use? Why? Can you rate your professional experiences in RS in terms of emotional support, work intensity, and stress level?}
\end{table*}

\end{document}